\documentclass[letterpaper,twocolumn,10pt]{article}
\usepackage{usenix-2020-09}

\usepackage{tikz}
\usepackage{amsmath}

\usepackage{filecontents}

\usepackage{xcolor}
\usepackage{booktabs}
\usepackage{amsthm}
\usepackage{multicol}
\usepackage{multirow}

\usepackage{color}   
\usepackage{hyperref}
\hypersetup{
    colorlinks=true, 
    linktoc=all,     
    linkcolor=blue,  
}

\newtheorem{prop}{Proposition}

\usepackage{enumitem}
\setlist[enumerate]{leftmargin=*}
\setlist[itemize]{leftmargin=*}

\begin{document}

\date{}

\title{A Certifiable Security Patch for Object Tracking in Self-Driving Systems via Historical Deviation Modeling}

\author{\rm{Xudong Pan, Qifan Xiao, Mi Zhang, Min Yang} \\ \textit{Fudan University, China} \\ 
\{xdpan18, 20210240056, mi\_zhang, m\_yang\}@fudan.edu.cn 
} 

\maketitle
\begin{abstract}
Self-driving cars (SDC) commonly implement the perception pipeline to detect the surrounding obstacles and track their moving trajectories, which lays the ground for the subsequent driving decision making process. Although the security of obstacle detection in SDC is intensively studied, not until very recently the attackers start to exploit the vulnerability of the tracking module. Compared with solely attacking the object detectors, this new attack strategy influences the driving decision more effectively with less attack budgets. However, little is known on whether the revealed vulnerability remains effective in end-to-end self-driving systems and, if so, how to mitigate the threat.

In this paper, we present the first systematic research on the security of object tracking in SDC. Through a comprehensive case study on the full perception pipeline of a popular open-sourced self-driving system, Baidu's Apollo, we prove the mainstream multi-object tracker (MOT) based on Kalman Filter (KF) is unsafe even with an enabled multi-sensor fusion mechanism. Our root cause analysis reveals, the vulnerability is innate to the design of KF-based MOT, which shall error-handle the prediction results from the object detectors yet the adopted KF algorithm is prone to trust the observation more when its deviation from the prediction is larger. To address this design flaw, we propose a simple yet effective security patch for KF-based MOT, the core of which is an adaptive strategy to balance the focus of KF on observations and predictions according to the anomaly index of the observation-prediction deviation, and has certified effectiveness against a generalized hijacking attack model. Extensive evaluation on $4$ KF-based existing MOT implementations (including 2D and 3D, academic and Apollo ones) validate the defense effectiveness and the trivial performance overhead of our approach.

\end{abstract}



\newcommand{\ego}[1]{\textcolor{blue}{#1}}

\maketitle

\section{Introduction} \label{sec:introduction}
\begin{figure}[t]
    \centering
    \includegraphics[width=0.5\textwidth]{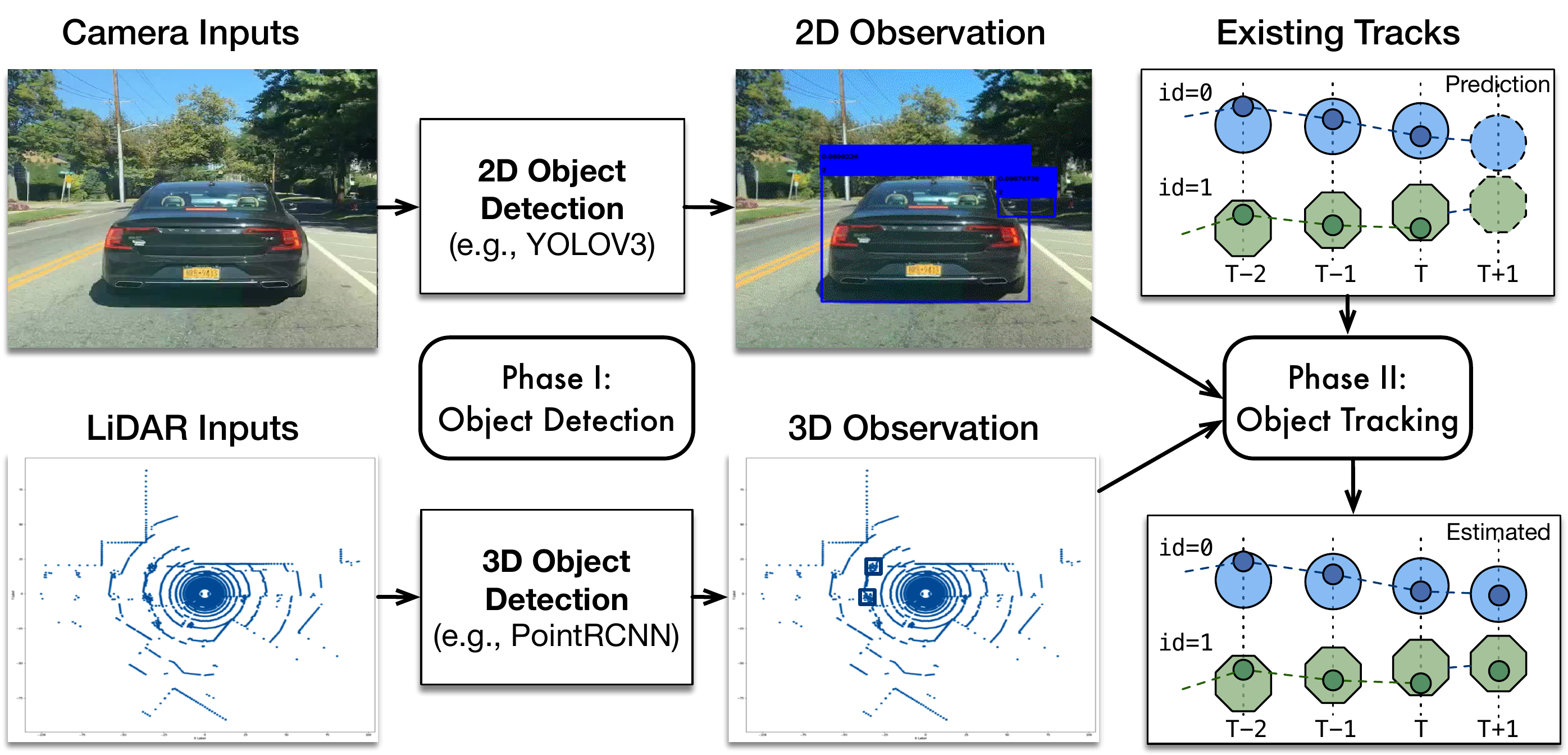}
    \caption{A typical perception pipeline in self-driving cars.}
    \label{fig:perception_pipeline}
\end{figure}

In the last few years, the development of self-driving cars (SDCs) is accelerating faster than ever \cite{BADUE2021113816}. To make proper and real-time driving decisions, commercial SDCs like Google's Waymo One \cite{Waymo} and Baidu's Apollo \cite{Apollo} implement the typical perception pipeline as in Fig.\ref{fig:perception_pipeline}: In the first half, the \textit{obstacle detection} module receives the environmental signals from multiple sensors (e.g., camera \cite{Redmon2018YOLOv3AI,Ren2015FasterRT,Liu2016SSDSS}, LiDAR \cite{Chen2019FastPR,Zhou2018VoxelNetEL,Liang2018DeepCF} and radar \cite{Sun2021WhoII}) and feeds them to the corresponding deep neural networks (DNNs) to detect the positions of the surrounding obstacles.

In the remaining half, the \textit{multi-object tracker} (MOT) builds the moving trajectories of the surrounding obstacles based on the detection results from the detection module \cite{Weng20203DMT,Kato2018AutowareOB,Feng2019MultiObjectTW,Yoon2016OnlineMT,Murray2017RealTimeMO}. Intuitively, the MOT matches all the detection results with the tracked trajectories of objects currently maintained in the module (i.e., \textit{data association}) and then update each track with the matched detection results with a predefined filtering algorithm (i.e., \textit{state estimation}). For example, Baidu's Apollo implements the Hungarian algorithm \cite{Kuhn1955TheHM} for data association and a per-trajectory Kalman Filter (KF \cite{Klmn1960ANA}) for state estimation. Finally, the updated trajectories are then sent to assist the subsequent prediction and planning pipeline to make decision on how the ego vehicle should behave to avoid collisions \cite{BADUE2021113816}.   

In the past years, the perception pipeline is intensively investigated for its security properties, especially for the first half. A majority of previous works present exploitation techniques to mislead the final detection results (e.g., appearing and disappearing) from the DNN-based object detectors (i.e., \textit{hijacking attacks})  \cite{Xie2017AdversarialEF,Chen2018RobustPA,Zhao2019SeeingIB,Cao2019AdversarialOA,Sun2020TowardsRL}. However, as pointed out in \cite{Jia2020FoolingDA}, most of previous attacks insufficiently consider the impact of the MOT module on the hijacked detection results. Designed for tolerating errors from the detection module, MOT implements a mature track management scheme to only allow the tracking results of detected objects with sufficient consistency and stability across multiple frames to be preserved and passed to the planning module \cite{Bernardin2008EvaluatingMO,Weng20203DMT}. This substantially inhibits the adversarial objects forged/hidden in few consecutive frames from actually affecting the ego vehicle's driving plan. 

Unfortunately, the MOT module itself is also exploitable. In \cite{Jia2020FoolingDA}, a new proof-of-concept (PoC) camera hijacking attack is composed against MOT, in which the attacker invokes bounding box shifting and disappearing attacks in sequence to let the MOT mistakenly perceive the front vehicle to hold a horizontal velocity. Propagating to the planning module, this mistake would result in a false turning left/right decision and may further cause crash accidents. However, this new PoC is specifically designed for a prototypical MOT system with camera inputs only. Key questions remain open on (i) whether the MOT module in end-to-end self-driving systems with MSF can still be exploited with the PoC and, (ii) if so, how to mitigate this severe safety threat.

To answer these open yet safety-critical questions, we present the first systematic study on the security of object tracking in self-driving systems. At the attack side, to achieve a comprehensive understanding on the vulnerability of existing KF-based MOT implementation in SDC, we conduct a case study on the full perception pipeline of Baidu's Apollo \cite{Apollo}, a popular open-sourced self-driving system where a KF-based MOT solution is implemented. We strikingly discover that, even when the multi-sensor fusion (MSF) of camera and LiDAR is enabled, an attacker is still able to mislead the ego vehicle to believe the front vehicle were moving horizontally even when the front vehicle is factually not. In contrast to the expected attack budgets of fooling the perception model of each sensor in the MSF, our constructed PoC only requires to manipulate the prediction results from the LiDAR-based perception module to achieve the same attack goal of \cite{Jia2020FoolingDA} on a real SDC. Such an attack budget is commonly recognized as feasible and practical.  

Rethinking the findings from our case study, we conclude that the mainstream KF-based MOT is inherently unsafe even with an enabled MSF mechanism. Through a careful root cause analysis, we reveal the vulnerability of the MOT module is mainly because, the adopted KF algorithm in state estimation contradicts with the general design principle of the tracking module. On the one hand, the KF algorithm fundamentally assumes the error from the observational data follows the Gaussian error model, which is unsuitable to deal with the heavy-tailed outlier data when the SDC is positioned under the adversarial environments. On the other hand, the updating formula in the KF algorithm is prone to emphasize more on the external observational data when it deviates more from the KF's own prediction. This characteristic contradicts with the MOT's role as an error-handling module for the noisy prediction results from the object detectors.

To address the design flaws, we propose a simple yet effective security patch for the mainstream KF-based MOT module. Instead of allowing the original KF algorithm to trust the external signal in an unconstrained manner, we propose an adaptive strategy which balances the contribution of the external observations and the internal predictions according to the anomaly index of the observation-prediction deviation from a robust estimator built upon historical deviation data. Via stochastic control theory, we prove our proposed security patch has certifiable mitigation effect against a generalized hijacking attack model, where the adversary can arbitrarily shift or hide the detection results in a fixed ratio of frames.

In summary, we mainly make the following contributions.
\begin{itemize}[topsep=4pt,itemsep=4pt,partopsep=4pt, parsep=2pt]
\item We systematize the attack taxonomy of MOT hijacking attacks (\S\ref{sec:security_setting}) and present a comprehensive case study to confirm the PoC in a prototypical 2D MOT \cite{Jia2020FoolingDA} also poses practical threats on Baidu's Apollo (\S\ref{sec:apollo_mot}-\ref{sec:confirmed_vul}).

\item We reveal the root vulnerability of existing KF-based MOT implementation in the inconsistency between the classical KF algorithm and its expected role in modern SDCs (\S\ref{sec:root_cause}).

\item We propose a simple yet certifiable security patch for the mainstream KF-based MOT module (\S\ref{sec:method_overview}-\ref{sec:method_modulation}), which is the first solution to mitigate the MOT hijacking attack and has provable effectiveness in lowering the trajectory deviation caused by potential attacks to a safe range (\S\ref{sec:security_analysis}).  

\item We conduct extensive evaluation with the KITTI dataset and $4$ mainstream 2D/3D MOT implementations from both the academy and the industry where our proposed security patch is installed. Empirical results validate the defense effectiveness of our approach on MOT hijacking attacks with almost no performance overhead on the normal MOT performance (\S\ref{sec:Experiment}).
\end{itemize}
\section{Background}
\subsection{Perception Pipeline}
In industry-level self-driving systems (e.g., Waymo \cite{Waymo} and Apollo \cite{Apollo}), the perception module plays the critical role in bridging environmental signals from multiple sensors (including camera, LiDAR, and radar) to the decision-making inner loop of an SDC (i.e., prediction-planning-control). In a typical processing frame, when the perception pipeline receives the 2D and 3D sensor inputs, a vector of deep learning models (e.g., YOLOV3 \cite{Redmon2018YOLOv3AI} for camera inputs and PointRCNN \cite{Shi2019PointRCNN3O} for LiDAR inputs) are employed to process the sensor inputs and detect the surrounding obstacles (e.g., other vehicles or pedestrians). Specifically, the surrounding obstacles are perceived as \textit{bounding box}es, a conventional notion in object detection in computer vision. For example, a 2D bounding box is characterized by two point coordinates, i.e., \textbf{bbox} $ = (x_1, y_1, x_2, y_2)$, which encodes the bottom-left and upper-right ends of the detection box. To determine more properties of the perceived obstacles, some contextual information (e.g., the class label) is also predicted along with the bounding box.   

The above procedure describes the first half of a typical perception pipeline, which produces a list of 2D and 3D bounding boxes for the surrounding obstacles (which may have missing/incorrect detection results). Then the rest of the pipeline, i.e., the object tracking module, builds the up-to-now trajectory of each surrounding obstacle, which lays an essential ground for the subsequent prediction (roughly, predicting the future trajectories of each obstacle based on the received up-to-now trajectories \cite{xu2020data}) and planning (roughly, planning the driving trajectory of the ego vehicle based on the predicted future trajectories \cite{Wan2022TooAT}). 


\subsection{Multiple Object Tracking}
\label{sec:prelim:mot}
The tracking module in the perception pipeline is a realization of multiple object tracking (MOT), a classical research topic of computer vision \cite{Luo2021MultipleOT} which aims at building the trajectories of multiple objects based on the given list of bounding boxes in each processing frame. At the $T$-th timestamp, each trajectory in an MOT module is identified with a surrounding object, and stores a sequence of its estimated \textit{driving state}s, i.e., $(s_1, \hdots, s_T)$, where each state $s_t$ usually describes the location (in the form of a bounding box) and the velocity of the object. For the simplicity of presentation, we consider the state definition of $s_t = (\widehat{\textbf{bbox}}_t, \vec{v}_t)$, where $\widehat{\textbf{bbox}}_t$ denotes the object bounding box and $\vec{v}_t$ denotes the velocity at each dimension. In the following, we consider a list of trajectories $\{(s_i(o))_{i=1}^{T}\}_{o\in\mathcal{O}_T}$ is maintained at the current timestamp (i.e., $\mathcal{O}_T$ is the set of identified objects).

When the MOT module receives the list of detected bounding boxes $\{\textbf{\text{bbox}}_i\}_{i=1}^{N}$ (i.e., \textit{observational data}) from the first half of the perception pipeline, the observational data goes through two main stages: \textit{data association} and \textit{state estimation}, which finally updates the maintained list of states to the latest version (i.e., $(s_i(o))_{i=1}^{T}\to(s_i(o))_{i=1}^{T+1}$). In the following, we mainly introduce the two stages in the context of Kalman filter (KF \cite{Klmn1960ANA}) based MOT algorithms, which are widely used in the industry-level self-driving systems (e.g., Baidu's Apollo \cite{Apollo}, Autoware \cite{Kato2018AutowareOB} and OpenPilot \cite{openpilot})

\noindent$\bullet$\textbf{ Data Association.} At the data association stage, the KF algorithm is first invoked on each trajectory to obtain the predict-only states at the $T+1$ timestamp. Formally, the predict-only states $\{s^{-}_{T+1}(o)\}_{o\in\mathcal{O}_T}$ are calculated by 
\begin{align}
    s^{-}_{T+1}(o) = As_{T}(o),
    \label{eq:kf_prediction}
\end{align} where $A$ is a \textit{state transition} operator specified with the assumed dynamic model. Then, a matching algorithm (e.g., the Hungarian algorithm \cite{Kuhn1955TheHM}) is applied to associate each trajectory with the optimal observational bounding box from   $\{\textbf{\text{bbox}}_i\}_{i=1}^{N}$ based on different similarity metrics between the observational bounding box and the one from the predicted state of the trajectory (i.e., the prediction data). For example, a common choice is to measure the Intersection over Union (IoU) between the observational and the prediction data \cite{xinshuo2020AB3DMOT}. A higher IoU means potentially better association. Ideally, data association aims at assigning the observational data to the identified object which physically generates the observational data. In this sense, data association is essential for the multiple object scenarios and determines the associated observational data for state updating. Formally, we denote the matching relation from the observational data to the objects as $\mathcal{M}:i\in\{1,2,\hdots,N\}\to\mathcal{O}_k\cup{}\{\text{null}\}$, where $\mathcal{M}(i) = \text{null}$ indicates that the observed bounding box $\textbf{bbox}_i$ matches no existing trajectories. 

\noindent$\bullet$\textbf{ State Estimation.} When a matching relation is established between an identified object $o$ with the predicted state $s^{-}_{T+1}(o)$ and the observed bounding box $\textbf{bbox}_i$. Then, MOT invokes the updating rule in the KF algorithm to merge the observational and the prediction data to obtain the final estimated state $s_{T+1}(o)$ at the next timestamp. Otherwise, if there is no matched observational data for the bounding box, the predicted state is immediately adopted as the final estimated state. Without loss of generality, we simplify the location representation of a bounding box as its center coordinate, which usually has independent estimators from the edge length of the bounding box. Below, denoting the center of the observational data $\textbf{bbox}_i$ as $z_{T+1}(o)$, we present the updating rule of the KF algorithm:
\begin{align}
    s_{T+1}(o) = s^{-}_{T+1}(o) + K_{T+1}(z_{T+1}(o) - Hs_{T+1}^{-}(o)),
    \label{eq:kf_merging}
\end{align} 
where $H$ is the \textit{state projection} operator, which intuitively projects the state vector $s_{T+1}^{-}$ to the space of the observational data, and $K_{T+1}$ is the well-known \textit{Kalman gain} operator, which is independently maintained for each trajectory and is updated by iteration based on the \textit{observation-prediction deviation} term $\delta_{T+1}(o) := z_{T+1}(o) - Hs_{T+1}^{-}(o)$. As Eq.(\ref{eq:kf_merging}) shows, the final estimated state $s_{T+1}$ can be viewed as an interpolation between the prediction ($s_{T+1}^{-}$) and the observational data ($z_{T+1}(o)$). As analyzed in \cite{Cipra1997KalmanFW}, an important property of the Kalman gain operator is that it is asympototic to $H^{-1}$ when $|\delta_{T+1}|$ increases, which \textit{lets the estimated state depend increasingly more on the observation}. Appendix \ref{sec:app:kalman_filter} presents a self-contained background material on the KF algorithm used in Baidu's Apollo.

\noindent$\bullet$\textbf{ Trajectory Management Scheme.}  With a static list of trajectories maintained in the system, the two stages above characterize one typical processing frame of MOT. When combined with the temporal dimension, the MOT module further adopts a trajectory management scheme to control the creation and the destroying of trajectories. Specifically, a new trajectory is created if an object is detected for over $H$ frames, while an existing trajectory is destroyed if there are $R$ consecutive frames where the trajectory is not matched with any bounding boxes. The hyperparameters $H$ and $R$ are commonly referred to as the \textit{hit count} and the \textit{reserved age}. 

From the perspective of system design, such a trajectory management scheme is mainly used for handling the possible prediction errors from the precedent object detection module. Either missing detection (i.e., \textit{false negatives}) or artifact detection (i.e., \textit{false positives}) in the detection results would have almost no effect on the death or birth of a trajectory, which makes the tracking results consistent over time. Moreover, according to \cite{Jia2020FoolingDA}, the trajectory management mechanism also intensively increases the attack budget of most existing appearing/disappearing attacks \cite{Xie2017AdversarialEF,Chen2018RobustPA,Zhao2019SeeingIB,Cao2019AdversarialOA,Sun2020TowardsRL} on self-driving systems. When considering the error-handling effects of the track management scheme, an attacker has to hide the target object away from the prediction results of the detection module in $R$ consecutive processing frames to cause its actual hiding in the perception of the MOT module \cite{Jia2020FoolingDA}. However, in practice, the reserved age $R$ is configured to be $60$ frames in a $30$ fps perception pipeline \cite{Zhu2018OnlineMT}, an almost infeasible number of consecutive successful hiding attacks existing attack techniques could achieve.

\section{Security Settings}
\label{sec:security_setting}



\subsection{Threat Model} To abstract the attack scenario in \cite{Jia2020FoolingDA} which targets at the camera-based perception pipeline, we consider a victim SDC which perceives the external environment with an abstract sensor $\mathcal{S}$ (e.g., a camera, a LiDAR, or both in a fused way).

\noindent$\bullet$\textbf{ Attack Goal.} An MOT hijacking attack wants to cause an unacceptable deviation between the victim SDC's perceived trajectory on the target object and its ground-truth trajectory. Formally, given the ground-truth driving trajectory of the target object $o_\text{tgt}$ up to the $T$-th timestamp is $(\hat{s}_t(o_\text{tgt}))_{t=1}^{T}$ and the perceived trajectory $({s}_t(o_\text{tgt}))_{t=1}^{T}$, the deviation at the $t$-th timestamp is measured by $d(\hat{s}_t, \hat{s}_t) = \|H\hat{s}_t(o_\text{tgt}) - Hs_t(o_\text{tgt})\|$, i.e., the Euclidean distance between the object centers in the two trajectories. By denoting the maximally tolerable deviation as $d_\text{max}$, we define an MOT hijacking attack succeeds if for some $t \in \{1, \dots, T\}$, we have $d(\hat{s}_t, \hat{s}_t) > d_\text{max}$, i.e., the perception deviation is over the secure threshold. The statistics in Table \ref{tab:standard_deviation} provides the actual $d_\text{max}$ to cause a successful \textit{off-road attack}, i.e., the front vehicle is perceived to be off the road, and \textit{wrong-way attack}, i.e., the front vehicle is perceived to be in an incorrect driving way, in both local and highway driving scenarios.

\begin{table}[t]
  \centering
  \caption{The required deviation in the trajectory perception (i.e., $d_\text{max}$) to cause real-world accidents \cite{junjie2020drift}.}
  \scalebox{0.65}{
    \begin{tabular}{cccc}
    \toprule
    Off-Road (Local) & Off-Road (Highway) & Wrong-Way (Local) & Wrong-Way (Highway) \\
    \midrule
    0.895m & 1.945m & 2.405m & 2.855m \\
    \bottomrule
    \end{tabular}}%
  \label{tab:standard_deviation}%
  \vspace{-0.25in}
\end{table}%


\noindent$\bullet$\textbf{ Security Assumptions.} \cite{Jia2020FoolingDA} mainly makes the following assumptions on the capability of the attacker.
\begin{itemize}[topsep=4pt,itemsep=4pt,partopsep=4pt, parsep=2pt]
    \item \textit{Assumption 1.} The adversary is able to perturb the received signals of $\mathcal{S}$ (within the physical capability of existing sensor hijacking techniques) in a $r\%$ ratio of detection frames when the target object is in the region-of-interest (RoI) of the ego car.
    \item \textit{Assumption 2.} The adversary has a white-box access to the object detection model with respect to the sensor $\mathcal{S}$.  
    \item \textit{Assumption 3.} The adversary has access to the current list of trajectories which are maintained in the MOT system and has identified the trajectory of the target object.
\end{itemize}
Below, we concisely analyze the rationale and the feasibility of the adopted assumptions above. \textit{Assumption 1} prepares the adversary with the attack surface, where he/she can maliciously perturb the visual inputs to the camera (e.g., by stickers \cite{Chen2018RobustPA}, posters \cite{Zhao2019SeeingIB}, and colored objects \cite{Cao2021InvisibleFB}) or the point cloud inputs to the LiDAR (e.g., by laser transmitter \cite{Cao2019AdversarialOA} and irregular 3D printing objects \cite{Cao2021InvisibleFB}) for tempting the misbehavior of the object detection modules. For example, in the first MOT hijacking attack, \cite{Jia2020FoolingDA} assumes the adversary may put an adversarial pixel patch to the camera inputs to conduct the exploitation. \textit{Assumption 2} is mainly adopted for efficiently generating adversarial perturbations related with certain abnormal model behaviors. In adversarial machine learning, the white-box access to the target model allows the adversary to leverage gradient-based optimization in searching for the optimal adversarial perturbation under the guidance of a differentiable attack utility function \cite{Szegedy2014IntriguingPO,Kurakin2017AdversarialEI,Carlini2017TowardsET}. In the context of SDC, the attacker can have access to the target model especially when the self-driving system is open-sourced. For example, it is easy to access the parameters and the architecture of the objection detection models in Baidu's Apollo, which are retrieved from the cloud server during the start-up process of the vehicle and are stored in the system's mounting disk. Finally, \textit{Assumption 3} allows the adversary to get information about the maintained trajectory of the target object in the victim SDC for devising the perturbation strategy on the observational bounding box of the target object, which we detail in the next section. However, the trajectory information is usually internal to the victim SDC, while \cite{Jia2020FoolingDA} pays no discussion to the feasibility of retrieving the information in practice. Our attack in Section \ref{sec:confirmed_vul} on the Apollo MOT module leverages a fuzzing approach to relax this assumption.


\subsection{Hijacking Attacks on 2D MOT}
\label{sec:2d_mot}
\noindent$\bullet$\textbf{ Attack Operations.} Under the security assumptions above, we introduce and discuss the hijacking attack on the camera-based MOT system in \cite{Jia2020FoolingDA}. In general, the PoC attack adopts two attack operations, namely, \textit{Shifting} and \textit{Hiding}, on the bounding box of the target object, which is usually chosen as the front vehicle of the ego vehicle to induce more severe consequences (e.g., emergent braking or crashing).   
\begin{itemize}[topsep=4pt,itemsep=4pt,partopsep=4pt, parsep=2pt]
    \item (i) \textit{Shifting}: The observed bounding box of the target object is shifted by $\delta_\text{atk}$, i.e., $\textbf{bbox}(o_\text{tgt}) \to \textbf{bbox}(o_\text{tgt}) + \delta_\text{atk}$.
    \item (ii) \textit{Hiding}: The bounding box of the victim is hidden from the perception of the SDC, i.e., $\{\textbf{bbox}_i\}_{i=1}^{N} \to \{\textbf{bbox}_i\}_{i=1}^{N}\setminus{\textbf{bbox}(o_\text{tgt})}$. 
\end{itemize}

According to previous studies \cite{Zhao2019SeeingIB,Cao2019AdversarialOA}, both hiding and shifting operations can be achieved by adding adversarial patches on the sensor input (under \textit{Assumption 1-2}). 

\noindent$\bullet$\textbf{ Attack Procedures.} To mislead the MOT module, the PoC attack devises a two-phase attack: (i) First, the attack exploits the trajectory updating mechanism to inject incorrect perception on the velocity of the front vehicle. Specifically, in the first attack frame, the attacker maliciously shifts the bounding box of the target object by $\delta_\text{atk} = \lambda v_\text{atk}$, where $v_\text{atk}$ is the direction of the artifact velocity specified by the attacker. When doing the shifting, the scale parameter $\lambda$ is determined based on the current list of maintained trajectories (\textit{Assumption 3}), under the constraint that the shifted bounding box would still be associated with the target trajectory. Otherwise, the malicious observational data would not update the target trajectory. After the shifted bounding box matches with the target trajectory, the updating rule of KF, i.e., Eq.(\ref{eq:kf_merging}), would put most of its weight on the observational data due to the deviation between its internal prediction $Hs_{T+1}^{-}(o_\text{tgt})$ and the perturbed observation $z_{T+1} + \lambda v_\text{atk}$ tends to be large. (ii) Then, the attack exploits the track prediction mechanism to preserve the ego vehicle's misconception on the velocity. Specifically, in the next several frames, the attacker maliciously hides the bounding box of the target object from the perception module. Therefore, according to Eq.(\ref{eq:kf_prediction}), the MOT module continues to adopt the predicted states of the target object as the next states in the hiding frames, until the trajectory is removed from the system after $R$ frames. As the next predicted states are very likely to maintain or even enlarge the artifact velocity under the predefined state transition operator $A$ (which is mainly designed for non-adversarial scenarios), the perceived trajectory would therefore accumulate the deviation along the direction $v_\text{tgt}$, which satisfies the attack goal and would lead to off-road or wrong-way consequences.

\section{From PoC to Threat on End-to-End System}
\label{sec:case_study}
\subsection{The Mechanism of Apollo-MOT}
\label{sec:apollo_mot}
To systematically understand the vulnerability of existing implementation of KF-based MOT in real-world SDC, we present the following case study on Baidu's Apollo, a popular open-sourced self-driving system. The Apollo system also implements a KF-based MOT module in its perception pipeline, which we refer to as \textit{Apollo-MOT} in the remainder of this paper. Different from the prototypical system studied in the PoC attack \cite{Jia2020FoolingDA}, Apollo-MOT implements a complicated fusion mechanism over the detection results from both the camera and the LiDAR. Besides, Apollo-MOT also involves more dimension in measuring the similarity between the observational bounding box and the predicted trajectory during the data association phase. 

\noindent\textbf{General Workflow.} In each detection frame, Apollo-MOT receives lists of 2D and 3D bounding boxes $\{\textbf{bbox}_i^{\text{2D}}\}_{i=1}^{N_1}$ and $\{\textbf{bbox}_i^{\text{3D}}\}_{i=1}^{N_2}$ from the camera-based and LiDAR-based object detectors respectively. Instead of building independent lists of trajectories for different sensors, Apollo-MOT maintains a universal list of trajectories each of which is identified with an obstacle in the RoI. Specifically, the behavior of an obstacle at a certain timestamp is characterized by a state containing both the 2D and 3D driving information (similar to the 2D case introduced in Section \ref{sec:prelim:mot}). Consequently, at the data association stage, Apollo-MOT matches the 2D and the 3D observational bounding boxes respectively with the current list of trajectories. Then, for the matched pairs of observation and trajectory, the corresponding part in the maintained object state is updated by the information from the same dimension. In this way, Apollo-MOT fuses the observational information from multiple sensors for more accurate and robust perception, but mainly in a non-adversarial setting \cite{Cao2021InvisibleFB}.

\begin{figure}[t]
    \centering
    \includegraphics[width=0.45\textwidth]{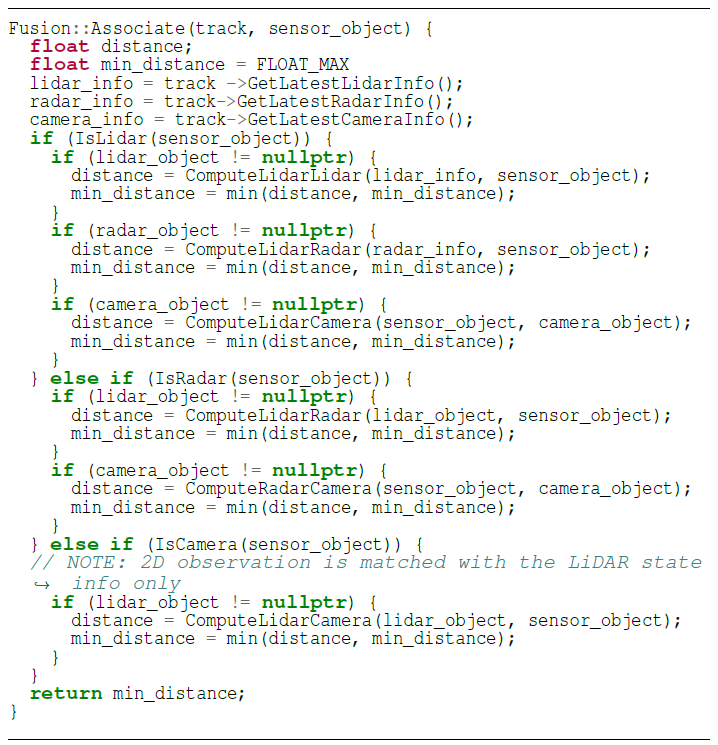}
    \caption{Simplified pseudocode of the MOT subsystem in the latest codebase of Apollo \cite{apollo_fusion}.}
    \label{fig:apollo_code}
\end{figure}

\begin{figure*}[t]
    \centering
    \includegraphics[width=1.0\textwidth]{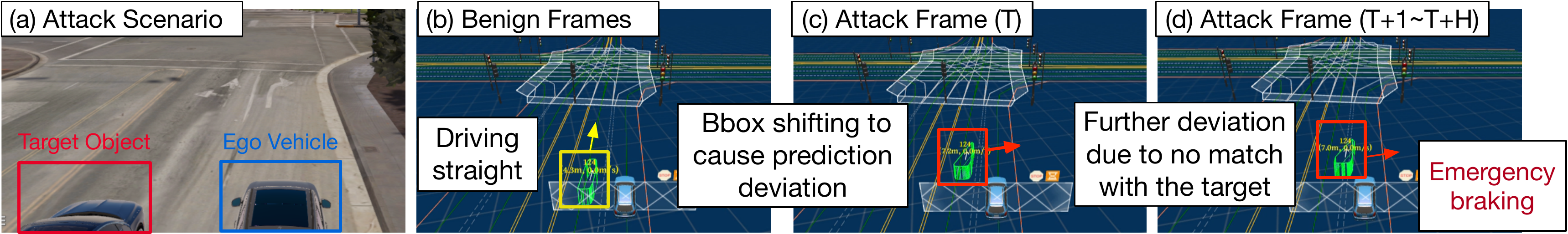}
    \caption{Simulator-based experiments which confirm the open-sourced Apollo self-driving system also suffers from the threat of MOT hijacking.}
    \label{fig:lgsvl}
\end{figure*}
\subsection{3D Information Dominates Apollo-MOT} 
By inspecting the design details of Apollo-MOT released in its latest codebase (v7.0.0 \cite{Apollo}), we observe that despite the above design of multi-sensor fusion (MSF), the 3D information still dominates the end-to-end workflow of Apollo-MOT, and also the downstream decision-making processes. This mainly reflects in the following aspects: (i) First, the 3D information is the main reference during data association. Instead of simply calculating the IoU between the observation and the prediction, Apollo-MOT calculates the similarity score in a more comprehensive way. On the one hand, more similarity dimensions (e.g., the shape, the direction, the feature) are involved in the calculation of the matching score. Especially for 3D bounding boxes, $7$ different similarity dimensions are adopted. On the other hand, Apollo-MOT introduces the inter-sensor matching score where the 3D observation is compared against both the 2D and 3D information in the predicted states for matching scores. Appendix \ref{sec:app:kalman_filter} provides more technical details on the data association part. (ii) Moreover, the matching result at the 3D part directly determines the birth/death of a trajectory. As the excerpted pseudocode in Listing \ref{lst:apollo_code} shows, the LiDAR information is the main reference in the fusion process. When a trajectory is matched with an observed 3D bounding box, Apollo-MOT marks the trajectory as matched (which means the number of consecutive missing frames would not accumulate) whenever there is a matched 2D bounding box. In contrast, even when the trajectory is matched with a 2D bounding box but not with a 3D one, the current matching status of the trajectory is still regarded as missing. In this sense, the matching status at the 3D part dominates the decision of whether increments or decrements the missing matching counter, which immediately determines the life cycle of the trajectory. (iii) According to the public codebase \cite{prediction_module} and a technical report by the Apollo team \cite{xu2020data}, the 3D information in the maintained trajectories of the surrounding obstacles plays a more important role in the downstream prediction and the planning modules.

In summary, even though Apollo implements the MSF mechanism in the MOT module, our analysis above reveals that by solely perturbing the sensor inputs to the LiDAR (no need of the camera), the attacker is still able to hijack the observed 3D bounding boxes (instead of the 2D bounding boxes) and thus the perceived trajectories of the target object, expecting more severe attack consequences in the meantime.

\subsection{Confirmed Vulnerability of Apollo-MOT}
\label{sec:confirmed_vul}
According to the feasibility analysis in Section \ref{sec:apollo_mot}, it is reasonable to expect that the MOT hijacking attack can realize the attack goal of deviating the perceived trajectory by perturbing the 3D point cloud. In the following, we report the empirical results which confirm the vulnerability on the end-to-end Apollo system running in the LGSVL simulator. Specifically, we conduct the MOT hijacking attack by simulating the driving scenario in Fig.\ref{fig:lgsvl}(a), where the target object drives straight by the side of the ego self-driving Apollo vehicle (Fig.\ref{fig:lgsvl}(b)). We conduct the similar attack as described in Section \ref{sec:2d_mot}, where the first attack frame happens at $T=6$. Instead of relying on the internal trajectory information of the victim Apollo car, we leverage the binary search on the perturbation scale $\lambda$ during the full driving process, with the objective of maximizing the trajectory deviation and of ensuring the object that its bounding box still matches with its corresponding trajectory in MOT in the first attack frame. We visualize the predicted trajectories of Apollo in Dreamview to inspect how the attack influences the driving decision of Apollo (Fig.\ref{fig:lgsvl}(c)-(d)). As Fig.\ref{fig:lgsvl}(d) shows, after the attack frames, Apollo-MOT would mispredict the trajectory of the target car, which is physically driving straight, to have an incorrect horizontal velocity  towards the ego car. This causes the emergency braking of the ego car, leading to a dangerous driving scenario. In practice, if with the supplement of physical apparatus (e.g., LASER emitter \cite{Cao2019AdversarialOA,Cao2021InvisibleFB}), it is feasible for the attackers to break the MSF-based Apollo-MOT via the MOT hijacking attack.

\subsection{Root Cause Analysis}
\label{sec:root_cause}
Finally, we discuss the root cause which underlies the confirmed vulnerability of the existing KF-based MOT implementation in self-driving systems. At the first glance, the dominance of the 3D information in Apollo-MOT should seemingly pay for the bill. However, we argue that this statement does not touch the essence of the vulnerability. On the one hand, it is a natural and reasonable choice for the perception module to rely more on the LiDAR information. According to the characteristics of the SDC-equipped sensors, LiDAR and 3D object detectors are usually more accurate in perceiving foreground obstacles than the 2D counterparts. Therefore, employing the 3D matching results as the main factor for track management would lead to less false positives/negatives. On the other hand, even if the MSF mechanism in Apollo-MOT were implemented in an alternative way of combining 2D and 3D detection results, the adversary may still be able to shift and hide the bounding boxes with existing attack techniques \cite{Cao2021InvisibleFB}. 

From our perspective, the cause of the vulnerability is mainly in the incompatibility between the classical KF algorithm and the role of MOT in the perception pipeline. As introduced in Section \ref{sec:2d_mot}, the KF algorithm is originally designed for a non-adversarial external environment (with an expected level of white noises), while the MOT module is designed for error-handling the prediction results from the detection module and generating consistent perception of the obstacle trajectories. When the MOT hijacking attack is not revealed by \cite{Jia2020FoolingDA}, three types of errors, i.e., false positives (artifact detection), false negatives (missing detection) and unbiased bounding box shifting, are already considered in the MOT design. The first is very likely to have no matched trajectory during data association and has almost no effect on the downstream processing, the second is handled by the mechanism of the reserved age and the default prediction-only rule in KF, while the final one is handled by the white noise in the dynamic model underlying KF (Appendix \ref{sec:app:kalman_filter}). However, when KF is first integrated into the self-driving systems, almost no consideration is devoted to handling a biased bounding box shift (which differs from the environmental noise in a Gaussian nature). As the KF algorithm is prone to trust more on the observational data when the observation-prediction deviation becomes larger, the trajectories in the KF-based MOT are easily hijacked by the maliciously shifted bounding boxes.

\section{A Certifiable Security Patch for KF-based MOT Implementations}
\label{sec:methodology}
\begin{figure}[t]
    \centering
    \includegraphics[width=0.45\textwidth]{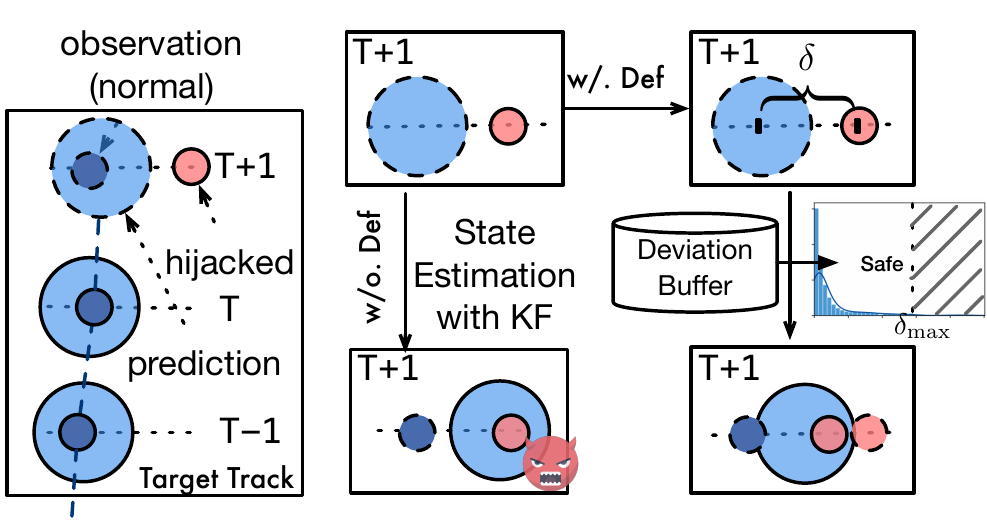}
    \caption{Overview of our methodology.}
    \label{fig:algorithm}
\end{figure}

\subsection{Methodology Overview}
\label{sec:method_overview}
Targeting at the root cause of the revealed vulnerability, our proposed security patch mainly strengthens the robustness of the existing KF-based MOT implementation in the following aspects:
(i) \textbf{Monitoring the Runtime Deviation}: We introduce an \textit{outlier-aware deviation buffer} $B$ which selectively records the observation-prediction deviation in each detection frame. This provides the decision basis to modulate the MOT updating process. 
(ii) \textbf{Adaptive Modulation on the Deviation Scale:} Instead of following the classical KF updating rule, we generalize the merging equation in Eq.(\ref{eq:kf_merging}) of the prediction $Hs_{T+1}^{-}$ and the observation $z_{T+1}$ into $s_{T+1} = \alpha s_{T+1}^{-} + K_{T+1,o}\phi_{T+1}(z_{T+1}(o) - Hs_{T+1}^{-})$, where $\phi$ is a modulation function derived adaptively based on the current deviation buffer. By design, the modulation function constrains the influence of the observational data on the final estimated state when the observation-prediction deviation is likely to be abnormal, while, when the deviation is normal to the historical records, the modulation function is silent. In Section \ref{sec:security_analysis}, we provide rigorous analytical results to prove the introduction of the modulation function would effectively control the deviation to be in the safe boundary when MOT is under the generalized hijacking attack, while causes almost no influence on the normal driving scenarios. An intuitive comparison between our proposed methodology and existing KF-based MOT can be found in Fig.\ref{fig:algorithm}.

\subsection{Monitoring the Runtime Deviation}
\begin{figure}[t]
    \centering
    \includegraphics[width=0.45\textwidth]{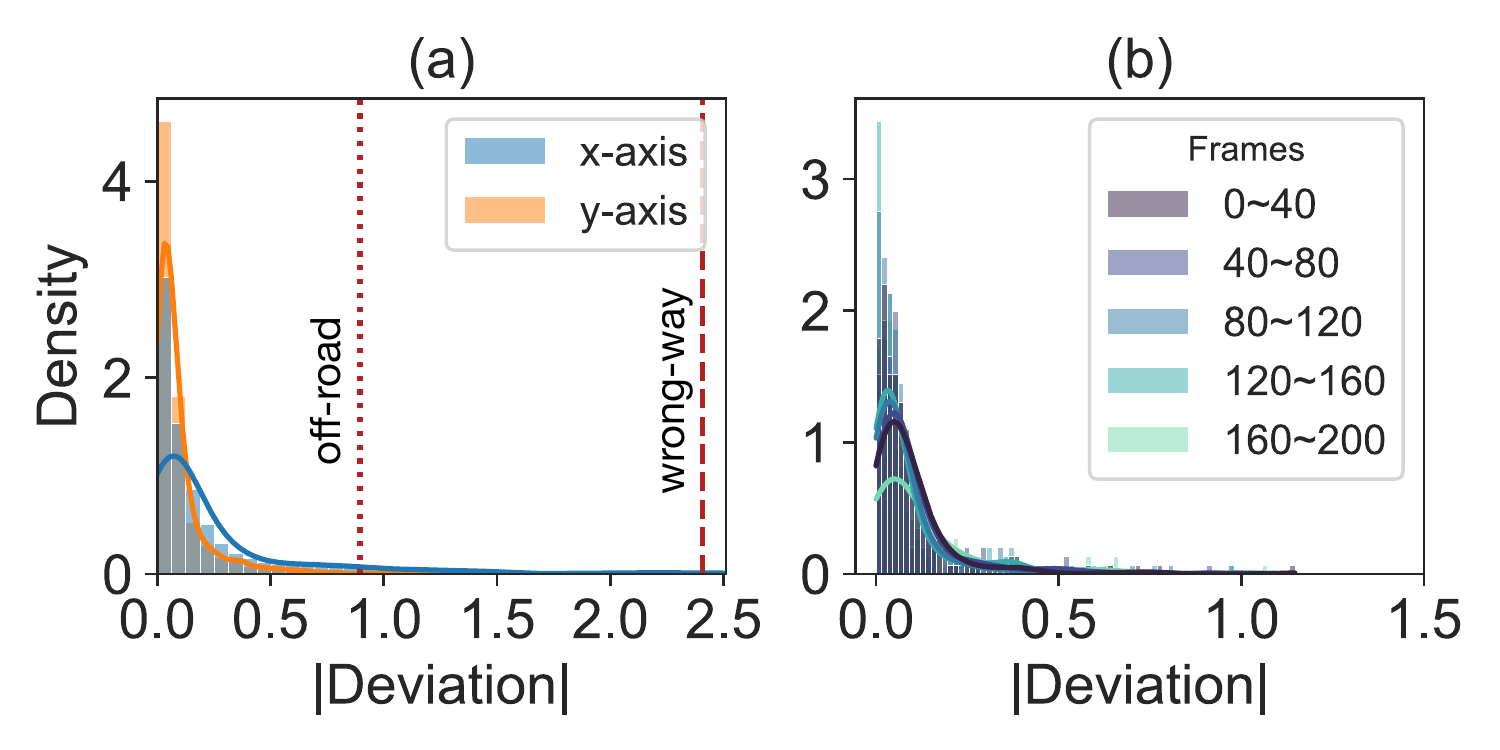}
    \caption{(a) Distribution of observation-prediction deviation at different axes and (b) in different time intervals at y-axis, where the red dotted vertical lines denote the minimum required deviation in a successful MOT hijacking attack.}
    \label{fig:demo_suite}
\end{figure}

To model the normal distribution of the observation-prediction deviation, we introduce a buffer $B_\Delta$ in the MOT runtime, which is shared for all the trajectories and is updated along the driving. In each processing frame, for every matched pair of observational data $z_{T+1}$ (i.e., the center of the observed bounding box) and the predicted state after projection $Hs_{T+1}^{-}$, we record the deviation vector $z_{T+1} - Hs_{T+1}^{-}$ into the buffer. Specifically, we highlight the following design choices for the self-driving context. 
\begin{itemize}
\item (i) \textit{Axis-Aware}:
According to an auxiliary statistical analysis on the ground-truth trajectories from the KITTI tracking dataset (Fig.\ref{fig:demo_suite}(a)), we confirm that, in consecutive frames, 
the deviation in different axes of the ego coordinate system has divergent distribution with one another.
Therefore, instead of calculating a comprehensive deviation metric (e.g., the Euclidean distance $\|z_{T+1} - Hs_{T+1}^{-}\|_2$), we propose to record the historical deviation in a per-axis way. For example, in the 3D scenario, we store the scalar deviation in respective buffers, i.e., each element in the deviation vector $z_{T+1} - Hs_{T+1}^{-}$.     
\item  (ii) \textit{Outlier-Aware}: In the adversarial setting, the attacker may already manipulate the bounding boxes to conduct the hijacking attack. As a result, depending on the raw deviation buffer for estimating the normal deviation distribution would be risky. To prevent the buffer poisoning, we propose to sanitize the deviation buffer after the collection in each frame, by eliminating the outlier values for each axis. Specifically, in our implementation, we choose the quantile-based outlier elimination scheme which only preserves the values in the $[\beta, 1-\beta]$ quantile of the deviation histogram,  where $\beta$ is a hyperparameter smaller than $50\%$.         

\item (iii) \textit{Time-Aware}: Moreover, considering the driving process of the ego vehicle dynamically changes in a short period of time, the normal distribution of the deviation would also face a drift after a certain time interval (Fig.\ref{fig:demo_suite}(b)). 
Therefore, we propose to cater for the timeliness of the driving scenarios by imposing a size constraint $S$ on the deviation buffer and implementing the First-in-First-Out (FIFO) buffer refreshing strategy. In this way, the recorded information in the deviation buffer updates smoothly with the gradual switch of the driving scenario, and would faithfully reflect the normal distribution in the current time interval.  
\end{itemize}

\subsection{Adaptive Modulation on the Deviation Scale}
With the design of the deviation buffer, our proposed security patch adaptively adjusts the influence of the observational data on the estimated state when abnormality is detected in the current observation-prediction deviation according to the recorded deviation statistics in the surrounding timestamps. In the following, we describe how to quantify the deviation abnormality and how to mitigate it. 


\noindent$\bullet$\textbf{ Quantifying the Deviation Abnormality.} 
As Fig.\ref{fig:demo_suite}(a) shows, the normal observation-prediction deviation scale mainly follows a light-tailed distribution. Statistically, the $\alpha$ index of the tail is close to $2$ for all axes ($1.86$, $2.30$, $2.51$ respectively), which indicates the similarity with a Gaussian tail \cite{Simsekli2019ATA}.
In other words, the distribution of the deviation scale has a trivial density on the $\{\delta > \delta_\text{max}\}$ (i.e., $\delta_\text{max}$, a sufficiently large threshold). This also conforms to the following physical interpretation: in a single detection frame (i.e., about $0.03$ seconds for a 30 \textit{fps} detection pipeline), a vehicle could hardly exhibit an intense horizontal shift due to the limit of the car's velocity. Therefore, the dynamic model underlying KF is able to stay with a low observation-deviation level in most normal driving cases. In contrast, the attacker in the MOT hijacking attack eventually aims at increasing the deviation of the predicted trajectory before and after the attack. \textit{Therefore, the MOT hijacking attack has almost no choice but to shift the bounding box over the normal threshold to maximize the deviation accumulation during the hiding stage.} Based on the above analysis, we propose to view the current observation-prediction deviation as abnormal when the value is larger than the $\alpha_\text{max} = 95\%$ quantile of the estimated Gamma distribution over the deviation buffer $B_{\Delta,T}$. Again, the above procedure is done in a per-axis way.

\noindent$\bullet$\textbf{ Modulate the Abnormal Deviation.}
\label{sec:method_modulation}
Denoting the $\alpha_\text{max}$-quantile threshold vector as $\delta_{\text{max}}$, we propose to modulate the raw deviation $\delta_T$ by the following coordinate-wise function 
\begin{equation}
\phi(\delta_{T, i}) = \begin{cases}
\text{sgn}(\delta_{T, i})\delta_{\text{max}, i}\quad\text{if }|\delta_{T, i}| > \delta_{\text{max}, i} \\ 
\delta_{\text{T}, i} \quad \text{Otherwise}
\end{cases}.
\end{equation}
In plain words, the modulation function $\phi$ clips the raw deviation to a scale no larger than the $\alpha_\text{max}$ quantile of the recorded deviation while preserves the direction of the deviation. Our following analytical results and experiments proves, such a simple security patch would indeed enhance the robustness of existing KF-based MOT implementation under hijacking attacks, while has almost no negative effect on the normal driving traces.

\subsection{Security Analysis}
\label{sec:security_analysis}
Finally, we present our analytical results regarding the security of the KF-based MOT system under a generalized hijacking attack, where the attacker adopts the shifting operations in $s\%$ ratio of frames and the hiding operations in an $h\%$ ratio of frames (s.t., $s\% + h\% \le r\%$, the total ratio of attack frames). Without loss of generality, we mainly study the deviation at the horizontal axis where the attacker targets, while similar deviation bound can be obtained for other axis. In the following results, we mainly establish the estimation on the trajectory deviation at the $T$-th frame when the SDC drives in different attack and defense configurations. We use $s_{ij}$ to denote the state variable in different scenarios, where $i, j\in\{0,1\}$ denote whether the observation is hijacked and whether the defense is deployed respectively. 

\begin{prop}[Attack Effectiveness] For the original KF algorithm, if the bounding box is shifted by $\lambda$ in $s\%$ ratio of frames (i.e., $\mathcal{T}_s$) and hidden in $h\%$ ratio of frames (i.e., $\mathcal{T}_h$), then the trajectory deviation grows with $T$ such that 
\begin{equation}
    \max_{\mathcal{T_s}, \mathcal{T}_h}|\Delta_{10}(T)| \sim \frac{\lambda}{\beta}\Theta(e^{-\beta s\% T})h\%T.
\end{equation}
where the maximal deviation is achieved when $\mathcal{T}_s = [(1-r\%)T, (1-h\%)T]$ and $\mathcal{T}_h = ((1-h\%)T, T]$. 
\label{prop:attack_effectiveness}
\end{prop}
The first proposition provides the theoretical justification on why MOT hijacking attack would succeed for the classical KF algorithm. In contrast, the situation becomes different when our security patch is installed. First, we prove the installed security patch has almost no influence on the MOT performance in the normal situation.

\begin{prop}[Trivial Defense Overhead] For the KF algorithm equipped with defense, the trajectory deviation $\Delta_{01}(t)$ is $0$ almost surely for every $t \ge 0$.
\label{prop:defense_overhead}
\end{prop}
In plain words, the proposition states that the adopted defense strategy would leave the estimated trajectory under defense unbiased from the one without defense, and therefore incurs almost no overhead on the normal performance of the MOT module. Finally, when under hijacking attacks, the following proposition proves our proposed security patch substantially mitigates the possible deviation.

\begin{prop}[Defense Effectiveness] For the KF algorithm equipped with defense, if the bounding box is shifted by $\lambda$ in $s\%$ ratio of frames and hidden in $h\%$ ratio of frames, then the trajectory deviation $\Delta_{11}(T)$ grows slower than the growth rate in Proposition \ref{prop:attack_effectiveness} by $\delta_\text{max}/\lambda$, i.e., 
\begin{equation}
    \max_{\mathcal{T_s}, \mathcal{T}_h}|\Delta_{11}(T)| \sim \frac{\delta_\text{max}}{\beta}\Theta(e^{-\beta s\% T})h\%T.
\end{equation}
\label{prop:defense_effectiveness}
\end{prop}
In summary, compared with the unbounded growth of trajectory deviation when no security patch is installed (cf. Proposition \ref{prop:attack_effectiveness}), Proposition \ref{prop:defense_effectiveness} implies that our defense indeed constrains the estimated trajectory deviation with a much slower deviation rates (as $\lambda \gg \delta_\text{max}$ when the attacker does want to cause a non-trivial deviation), while, according to Proposition \ref{prop:defense_overhead}, the installed security patch would have almost no negative effect on the normal MOT prediction. Once the hyperpameters $\delta_\text{max}$ are properly chosen, the constant bound is provably under the minimal required deviation to reach a successful attack as in Table \ref{tab:standard_deviation}. Appendix \ref{sec:app:proofs} provides the omitted proofs for the analytical results above.

\section{Evaluation and Analysis} \label{sec:Experiment}
In this section, we evaluate the performance and the attack resilience of $4$ mainstream 2D/3D MOT implementations under different attack/defense configurations with the object tracking section of the KITTI  \cite{andreas2012kitti} dataset, a popular benchmark dataset which is widely used for validating novel attack and defense insights for self-driving systems (e.g., \cite{Cao2021InvisibleFB,Hau2021ShadowCatcherLI,Zhu2021CanWU}). 

\subsection{Evaluation Setups}\label{sec:Experiment:setup}
\noindent$\bullet$\textbf{ Target MOT Implementations.} We choose two mainstream KF-based MOT implementations for the 2D and 3D tracking tasks respectively. For the 2D tracking task, we choose the implemented MOT module in the official implementation of \cite{Jia2020FoolingDA} (i.e., \textit{Jia's 2DMOT}), whose mechanism is described in Section \ref{sec:prelim:mot}, and the camera-based MOT implementation in Apollo (i.e., \textit{Apollo} 2DMOT). For the 3D tracking task, we choose \textit{AB3DMOT} \cite{xinshuo2020AB3DMOT} and LiDAR-based MOT module in Apollo (i.e., \textit{Apollo} 3DMOT).

\noindent$\bullet$\textbf{ Performance Metrics.} We measure the normal MOT performance with seven standard metrics from the CLEAR evaluation protocol \cite{keni2008CLEAR} ($\uparrow/\downarrow$ denotes the higher/lower the better):

    \noindent{\textit{(1) Multi Object Tracking Precision (MOTP $\uparrow$)}} mainly measures the performance of MOT on estimating the locations of tracking objects. Formally, MOTP is defined as:
\begin{align}
MOTP=\frac{\sum_{i,t}d^i_t}{\sum_tc_t},
\end{align}
where $d^i_t$ refers to the center distance of the $i$-th matched trajectory-object pairs predicted by MOT in the $t$-th frame, and $c_t$ refers to the total matched trajectory-object pairs predicted by MOT in the $t$-th frame.
    
    \noindent{\textit{(2) Multi Object Tracking Accuracy (MOTA $\uparrow$)}} mainly evaluates the performance of MOT on matching the objects with the trajectories. Formally, MOTP is defined as:
\begin{align}
MOTA=1-\frac{\sum_t(m_t+{fp}_t+{mme}_t)}{\sum_tg_t},
\end{align}
where $m_t$ refers to the number of unmatched objects predicted by MOT in the $t$-th frame, ${fp}_t$ refers to the number of unmatched trajectories in the $t$-th frame, ${mme}_t$ refers to the count of mistakenly matched trajectory-object pairs in the $t$-th frame, and $g_t$ refers to the total matched trajectory-object pairs in the ground truth of the $t$-th frame. 


\noindent{\textit{(3) Precision ($\uparrow$), (4) Recall ($\uparrow$) and (5) F1 ($\uparrow$)}}: In the contexts of MOT, precision measures the ratio of the correct matched trajectory-object pairs over all the matched trajectory-object pairs maintained by MOT, while recall is the ratio of the matched trajectory-object pairs successfully predicted by MOT over all the trajectory-object pairs in the ground truth. F1 is derived conventionally from precision and recall.
    
    \noindent{\textit{(6) Most Tracked (MT $\uparrow$) and (7) Mostly Lost (ML $\downarrow$)}:} Specifically, MT measures the ratio of ground-truth trajectories which are covered by a single trajectory maintained by MOT for at least $80\%$ of their each life span, while ML measures the ratio of ground-truth trajectories which are not covered by any trajectory maintained by MOT for more than $20\%$ of their each life span.
    
 On the other hand, we measure the resilience of the MOT implementation with two adversarial metrics:    
    
\noindent{\textit{(1) False Deviation (FD)}} is the deviation of target object's center positions on the attacker-specific axis. It represents the directly negative influence of hijacking attack on MOT. Besides, with the knowledge of the lane's width in the real world, we can further analyze the possibility of MOT-related accidents caused by hijacking attack. Specifically, the deviations in 2D MOT are measured in pixels, while the deviations in 3D MOT are measured in meters.
    
\noindent{\textit{(2) Lost Frames (LF)}} is the increment of  the unmatched times of target trajectories maintained by the MOT. It represents the duration of negative influence remained in the target MOT.

\noindent$\bullet$\textbf{ Dataset Statistics.} We adopt the tracking data in KITTI as our testing set. Specifically, there are $21$ real-world traces in KITTI, each of which consists of a series of continuous results of detection and the corresponding ground truth of MOT in each frame. For the 2D MOT, there are $4$ numerical data in pixels to represent the center position and the size of the 2D bounding box of the object. For the 3D MOT, there are $7$ numerical data in meters to represent the center position, the size and the orientation of the 3D bounding box of the object.


\noindent$\bullet$\textbf{ Attack Settings.} In the 2D case, we directly deploy the hijacking attack \cite{Jia2020FoolingDA} on the target MOT. In the 3D case, we follow the hijacking idea in the 2D case. First, we choose the horizontal axis of the ego car, which is also the left-right direction for the driving direction, as the target axis and select the car which is moving in the same or opposite direction as the target object. Next, we use the binary search to determine the optimal perturbation scale to maximize the deviation in the target axis, while ensuring the object that its bounding box still matches with its corresponding trajectory in MOT in the first attack frame. Finally, we hide the bounding box for at most $5$ frames to ensure the bounding box of this object will not longer match with its original trajectory in MOT for deviation accumulation.

\subsection{Mitigation Effects against MOT Hijacking}\label{sec:Experiment:attack}
First, we measure the robustness of different KF-based MOT impementations under hijacking attacks. Table \ref{tab:2D_attack} \& \ref{tab:3D_attack} respectively show the adversarial metrics of all the $4$ MOT implementations when our proposed security patch is installed or not (i.e., w/. and w/o. Defense). To better understand the real-world threats caused by hijacking attack, we also compare the reported FD metric with the minimum required deviation in Table \ref{tab:standard_deviation} to validate our proposed security patch reduces the false deviation to the safe range.

\begin{table}[t]
  \centering
  \caption{The influence of hijacking attacks on 2D MOT with and without our security patch.}
  \scalebox{0.75}{
    \begin{tabular}{lrrrr}
    \toprule
          & \multicolumn{2}{c}{\textbf{Jia's 2DMOT}} & \multicolumn{2}{c}{\textbf{Apollo 2DMOT}} \\
    \cmidrule(lr){2-3}    \cmidrule(lr){4-5}
          & \multicolumn{1}{c}{w/o. Defense} & \multicolumn{1}{c}{w/. Defense} & \multicolumn{1}{c}{w/o. Defense} & \multicolumn{1}{c}{w/. Defense} \\
    \midrule
    FD (max) & 413   & 196 ($2.11\times\downarrow$)   & 126  & 91 ($1.38\times\downarrow$) \\
    FD (avg) & 30.85 & 25.69 ($1.20\times\downarrow$)  & 37.20 & 27.05 ($1.38\times\downarrow$)  \\
    \midrule
    LF (max) & 750   & 730 & 18 & 7 \\
    LF (avg) & 237.62  & 248.19  & 2.60 & 1.05 \\
    \bottomrule
    \end{tabular}}%
  \label{tab:2D_attack}%
\end{table}%



\begin{table}[t]
  \centering
  \caption{The influence of hijacking attacks on 3D MOT with and without our security patch.}
  \scalebox{0.75}{
    \begin{tabular}{lrrrr}
    \toprule
          & \multicolumn{2}{c}{\textbf{AB3DMOT}} & \multicolumn{2}{c}{\textbf{Apollo 3DMOT}} \\
    \cmidrule(lr){2-3} \cmidrule(lr){4-5}
          & \multicolumn{1}{c}{w/o. Defense} & \multicolumn{1}{c}{w/. Defense} & \multicolumn{1}{c}{w/o. Defense} & \multicolumn{1}{c}{w/. Defense} \\
    \midrule
    FD (max) & 1.71  & 0.58 ($\mathbf{2.95}\times \downarrow$)  & 4.53  & 0.3 ($\mathbf{15.1}\times \downarrow$) \\
    FD (avg) & 0.27  & 0.09 ($\mathbf{3.00}\times \downarrow$) & 0.79  & 0.05 ($\mathbf{15.8}\times \downarrow$) \\
    \midrule
    LF (max) & 367   & 367   & 2     & 2 \\
    LF (avg) & 49.00  & 47.52  & 0.10  & 0.10  \\
    \bottomrule
    \end{tabular}}%
  \label{tab:3D_attack}%
\end{table}%




\noindent\textbf{Results \& Analysis.} As we can see from Table \ref{tab:3D_attack}, our security patch effectively mitigates the malicious deviation caused by hijacking attacks. For example, the maximal FD of AB3DMOT is reduced by $2.95\times$ and the average FD is reduced by $3.00\times$, while the maximal FD of Apollo 3DMOT is reduced by $15.1\times$. According to the minimum required deviation of off-road attack and wrong-way attack shown in the Table \ref{tab:standard_deviation}, in most of the test cases the malicious false deviation is mitigated to the safe region ($0.58, 0.3 < 0.895$, the minimum required deviation to cause an off-roard attack in a local road). Therefore, the ASR of hijacking attack are both decreased to $0.00\%$ after deploying our security patch on the 3D MOT implementations, which implies that our security patch does help existing 3D MOT implementation more robust to MOT hijacking attacks.

\begin{figure*}[t]
    \centering
    \includegraphics[width=1.0\textwidth]{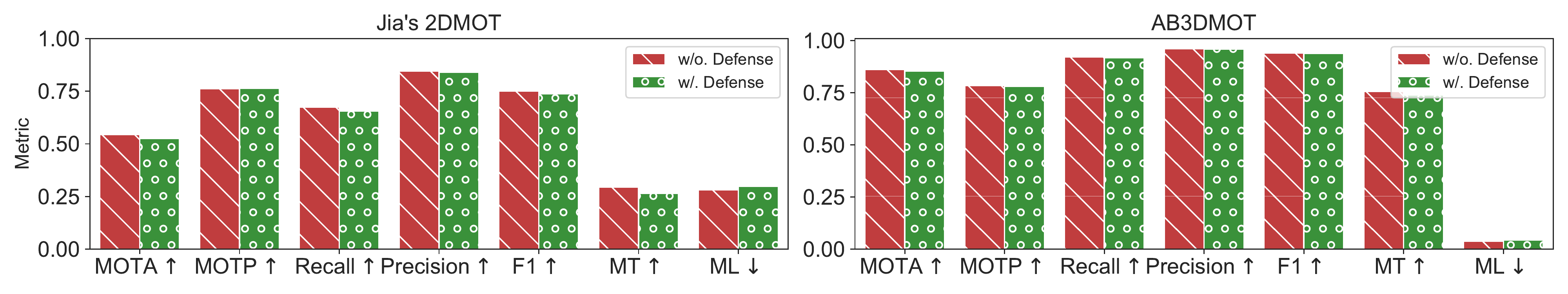}
    \caption{The performance of 2D and 3D MOT with and without our security patch on clean samples.}
    \label{fig:2D+3D_normal}
\end{figure*}

\begin{figure}[h]
    \centering
    \includegraphics[width=0.5\textwidth]{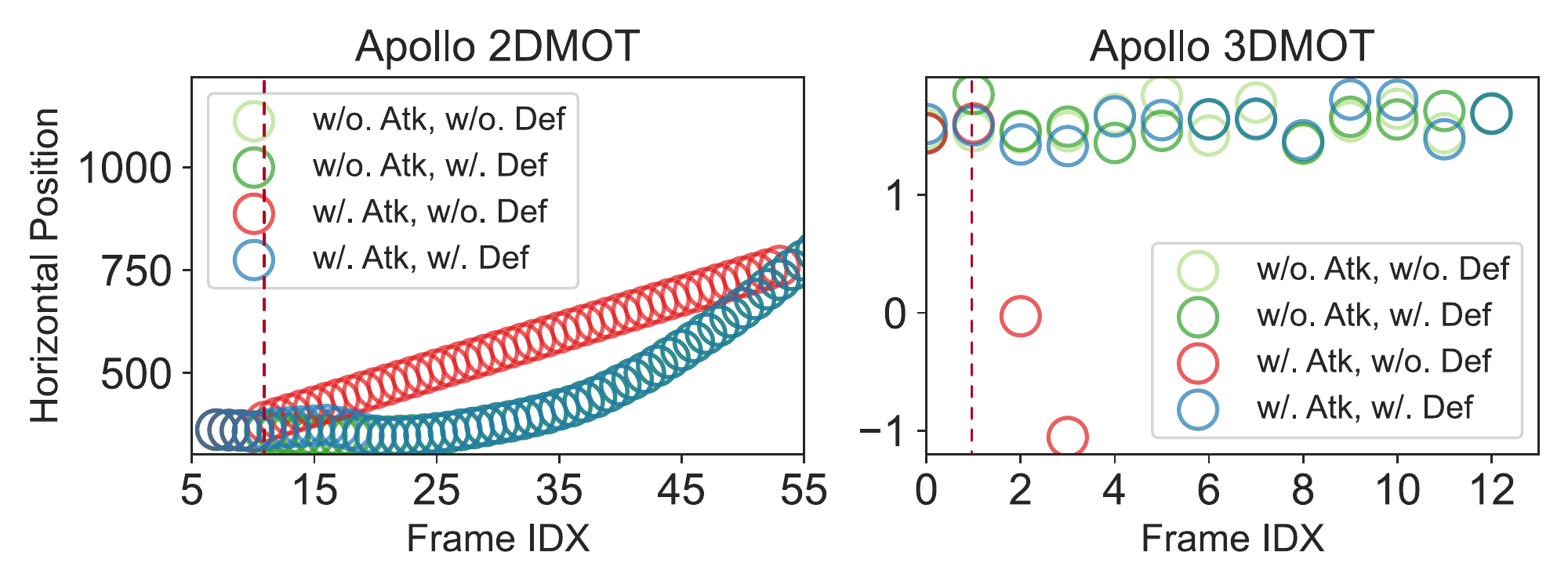}
    \caption{Visualization of the trajectories of the target object under different attack/defense configurations, where the red dashed lines mark the attack start frames.}
    \label{fig:visual:2d_mot}
\end{figure}

As is shown, the threats of hijacking attacks are more severe in the 2D scenario. As Table \ref{tab:2D_attack} shows, the FD caused by hijacking attacks can reach $413$ pixels at most, which is almost half the width of the input image of the perception module. Nevertheless, the FD of 2D MOT is also mitigated after implementing our security patch. For example, the average FD of Jia's 2DMOT is reduced by $2.11\times$, while the average FD of Apollo 2DMOT is reduced by $1.20\times$. For better intuition on the mitigation effect, Fig.\ref{fig:visual:2d_mot} visualizes the trajectories of the target object from 2D/3D cases randomly sampled from the KITTI dataset. As is shown, with our patch, the predicted trajectories from the Apollo MOT implementation is almost not affected by the malicious attacks 

In summary, our security patch effectively reduces the false deviation caused by hijacking attacks and guarantee the accuracy of the MOT prediction in adversarial settings. However, our security patch is not designed for preventing other negative effects such as ID switch, which may be harmless to the further planning of self-driving system, but would affect the MOT performance measured in CLEAR metrics \cite{keni2008CLEAR}.

\subsection{Overhead on Normal Performance}\label{sec:Experiment:normal}
Next, we evaluate the performance of KF-based MOT in the normal circumstances. However, to correctly evaluate the performance of Apollo MOT requires some technical hyperparameters which are invisible to us, we mainly report the clean performance of two non-Apollo MOT in this part. Fig.\ref{fig:2D+3D_normal} reports the performance of Jia's 2DMOT \cite{Jia2020FoolingDA} and AB3DMOT \cite{Weng20203DMT} on tracking cars.



\noindent\textbf{Results \& Analysis.} As Fig.\ref{fig:2D+3D_normal} shows, our security patch has almost no performance overhead on the normal processing of 3D MOT in the all $7$ evaluation metrics. For example, the MOTA and MOTP of AB3DMOT equipped with our security patch only decrease by less than $0.01$ ($1$\textperthousand~relatively), while the recall, precision and F1 metrics of AB3DMOT when our security patch is installed only decrease less than $0.02$ ($2\%$ relatively). Similar results are also observed in the 2D MOT's part of Fig.\ref{fig:2D+3D_normal}, the degradation of precision, recall and F1 of Jia's 2DMOT equipped with our security patch is also controlled below $0.01$. Besides, an increased MOTP is observed in some cases when evaluating Jia's 2DMOT equipped with our security patch, which indicates our security patch may also benefit the normal performance to some extension. In summary, our security patch incurs almost no performance overhead on the behavior of the 2D and 3D MOT implementations in non-adversarial settings.

\subsection{Ablation Studies}\label{sec:Experiment:ablation}
\noindent\textbf{Experiment Settings.} Furthermore, to explore the impact of different designs components on our security patch, we evaluate consider the following five comparison groups, namely, \textbf{A.} \textit{Gaussian}: Instead of Gamma distribution, our security patch uses the Gaussian distribution to model the normal deviation distribution. \textbf{B.} \textit{Elimination}: Instead of recording the deviation values from all the axes, we only store the largest deviation from the three axes into the buffer. \textbf{C.} \textit{Outlier-Unaware}: We do not use the outlier elimination mechanism in our full design. \textbf{D.} \textit{Axis-Unaware}: Instead of modeling the axis per axis, we model the deviation from all axes as a unified distribution. \textbf{E.} \textit{Axis-Unaware}: With the full design in Section  \ref{sec:methodology}, we vary the hyperparamter $\alpha_\text{max}$ in our security patch as other values: $0.8, 0.85, 0.99$. Table \ref{tab:delta} reports the performance of our security patch on different comparison groups on 3D MOT implementations, while Fig.\ref{fig:visual:2d_mot} reports the performance of our security patch with different values of $\alpha_\text{max}$ on 3D MOT.

\begin{figure}[t]
    \centering
    \includegraphics[width=0.5\textwidth]{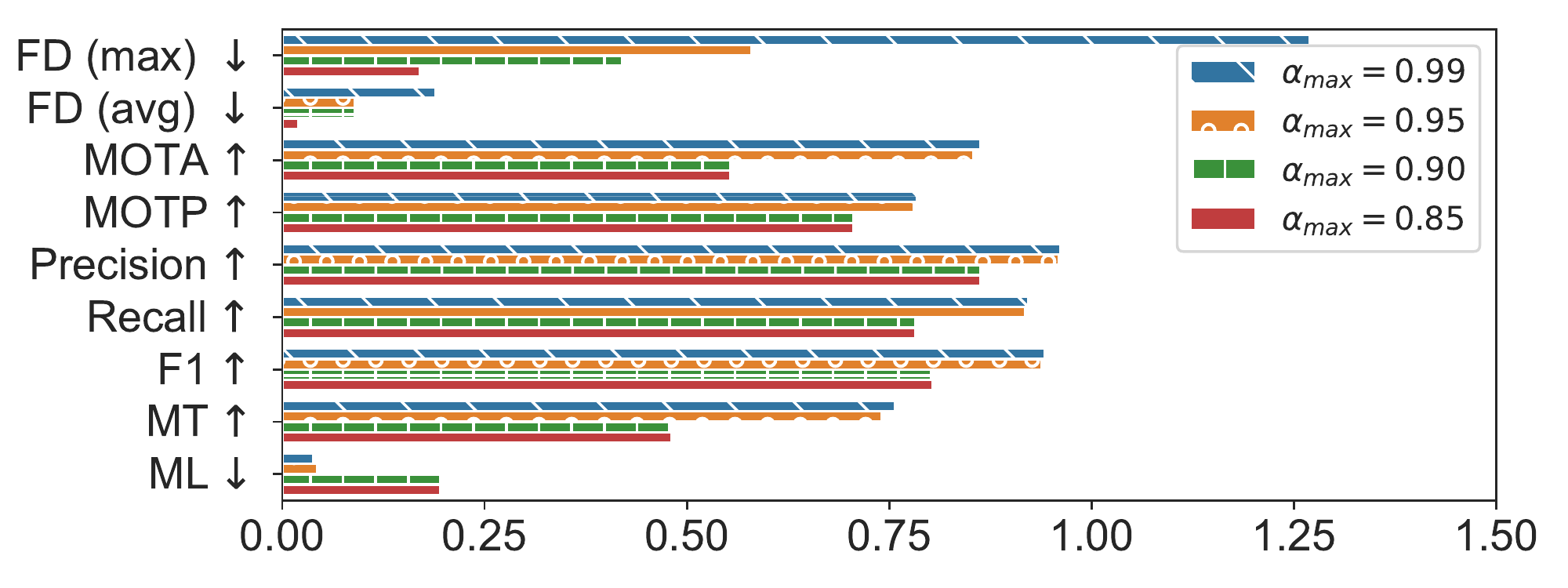}
    \caption{Mitigation effectiveness and performance overhead of our security patch on 3DABMOT when the hyper-parameter $\delta_\text{max}$ is chosen as the $\alpha_\text{max}$ quantile of the historical deviation distribution.}
    \label{fig:delta}
\end{figure}

\noindent\textbf{Results \& Analysis.} As we can see from Table \ref{tab:delta}, on the one hand, the Elimination, Outlier-Unaware and Axis-Unaware will also preserve the well performance of 3D MOT on CLEAR metrics, and Elimination and Outlier-Unaware even performs better than our original security patch. However, the robustness with these three methods of 3D MOT are significantly reduced. For Outlier-Unaware and Axis-Unaware, the average FD in average are $186.8\times$ larger than our original security patch on Apollo 3DMOT. For Elimination, its maximal FD is $1.95\times$ larger, and the average FD is $1.11\times$ larger than our original security patch on Apollo 3DMOT. On the other hand, although Gaussian performs well against hijacking attack, its performance on CLEAR metrics is a total disaster: its MOTA is only $34.20\%$ of our original security patch on AB3DMOT, and its F1 is only $71.11\%$ of our original security patch on AB3DMOT.


Moreover, as Fig. \ref{fig:delta} shows, the different values of $\alpha_{max}$ also significantly affect the performance of our security patch, while our final choice of $\alpha_{max} = 95\%$ is the empirical optimal choice. For example, if we choose a larger $\alpha_{max}$ (e.g., $\alpha_{max}=0.99$), the robustness of AB3DMOT degrades. For example, the maximal and the average FD increases by $1.18\times$ and $1.11\times$. Otherwise, if we choose a smaller $\alpha_{max}$ (e.g., $\alpha_{max}=0.9$ or $\alpha_{max}=0.85$), the normal performance of AB3DMOT will be harmed, e.g., the MOTA is reduced by at least $35.13\%$ and its F1 is reduced by at least $14.29\%$. In summary, the reported results in this part empirically validates the current design features of our security patch is indispensable to achieve the mitigation effectiveness against MOT hijacking while bringing almost no performance overhead.

\begin{table}[t]
  \centering
  \caption{The performance of 3D MOT when our security patch is installed with other alternative designs and other varied hyperparameters.}
  \scalebox{0.6}{
    \begin{tabular}{lrrrrrr}
    \toprule
          & \multicolumn{4}{c}{\textbf{AB3DMOT}} & \multicolumn{2}{c}{\textbf{Apollo MOT (3D)}} \\
    \cmidrule(lr){2-5} \cmidrule(lr){6-7}
          & \multicolumn{1}{c}{FD (max)} & \multicolumn{1}{c}{FD (avg)} &  \multicolumn{1}{c}{MOTA $\uparrow$} & \multicolumn{1}{c}{F1 $\uparrow$} & \multicolumn{1}{c}{FD (max)} & \multicolumn{1}{c}{FD (avg)} \\
    \midrule
    \textbf{Gaussian} & 0.69 & 0.07  & 0.292 & 0.667 & 0.15 & 0.02   \\
    \textbf{Elimination} & 1.71 & 0.19  & 0.864 & 0.943 & 7.36 & 0.75   \\
    \textbf{Outlier-Unaware} & 0.65 & 0.09  & 0.862 & 0.941 & 55.60 & 9.37 \\
    \textbf{Axis-Unaware} & 1.27 & 0.12  & 0.802 & 0.914 & 55.60 & 9.39 \\
    \midrule
    \textbf{Full Design} ($\alpha_\text{max} = 0.95$) & 0.58 & 0.09 & 0.854 & 0.938 & 0.30 & 0.05 \\
    \midrule
    \textbf{Full Design} ($\alpha_\text{max} = 0.99$) & 1.27 & 0.19 &  0.863 & 0.942 & 0.96 & 0.16 \\
    \textbf{Full Design} ($\alpha_\text{max} = 0.90$) & 0.42 & 0.17 &  0.554 & 0.804 & 0.30 & 0.06  \\
    \textbf{Full Design} ($\alpha_\text{max} = 0.85$) & 0.17 & 0.02 &  0.554 & 0.804 & 4.37 & 0.75  \\
    \bottomrule \end{tabular}}%
  \label{tab:delta}%
\end{table}%

\section{Discussion and Future Works}
\noindent$\bullet$\textbf{ On the Effect of False Alarms.} Generally, our proposed security patch works by modeling the normal observation-prediction deviation in the recent driving history in order to constrain the influence of the current observation when being a distributional outlier. As validated by the experimental results, we control the outlier threshold $\delta_\text{max}$ as $95\%$-quantile to empirically minimize the possible negative effect on the normal MOT performance, which may be further tuned in simulation environments when the security patch is to be deployed in real-world systems. 

Yet, we admit that there would be rare cases in non-adversarial settings that the observation-prediction deviation may be over the outlier threshold. Especially, when the target vehicle experiences extreme driving scenarios (e.g., a car accident ahead or its front vehicle stops abruptly), the target vehicle may abruptly steer the direction, resulting in an abnormal horizontal deviation. In the above cases, the underlying dynamic model in existing KF algorithms could hardly predict such extreme events and therefore would result in the deviation scale larger than most of the normal cases. 

From our perspective, although our proposed methodology would view the deviation as outliers and clip the scale to $\delta_\text{max}$, we argue that the operation would hardly influence the driving behaviors compared with the case when our security patch is not deployed. In fact, in contrast to the adversarial setting where the observation on the target vehicle is hidden in the next consecutive frames, the SDC would continuously observe the bounding boxes of the target vehicle, which helps adjust the KF algorithm's modeling on the target vehicle. On the one hand, if the abnormal deviation happens in only single frames, the SDC would very likely to observe normal observation in the next frame and update the state of the target vehicle in the classical KF way. On the other hand, if the abnormal driving pattern happens in several consecutive frames, the KF algorithm would adapt to the upcoming observational data, make more accurate prediction and finally result in a non-outlier observation-prediction deviation. In summary, the limited number of false alarms on the rare outlier deviation from non-adversarial driving scenarios would have almost no effect on the MOT's performance, conforming to our evaluation results in Section \ref{sec:Experiment:normal}.
 
 \noindent$\bullet$\textbf{ On the Coverage of Our Study.} Our current work mainly focuses on enhancing the robustness of the 2D and 3D KF-based MOT algorithms when they are installed with the proposed security patch. Such coverage is meaningful because the KF-based MOT algorithms are widely implemented in industry-grade self-driving systems, including Baidu's Apollo \cite{Apollo}, Autoware \cite{Kato2018AutowareOB} and OpenPilot \cite{openpilot}. Unfortunately, with less consideration on potential adversarial attacks, existing implementations of KF-based MOT in SDC still adopt the classical KF algorithm which adopts the assumption of Gaussian observational noise. Consequently, they are proved to be vulnerable against malicious manipulation on the observed bounding boxes, due to their over-confidence on adopting the seemingly abnormal observation as if it were true (\S\ref{sec:case_study}). 
 
 Despite the existence of few mathematical works which aims at enhancing the capability of the classical KF in handling outliers \cite{Cipra1997KalmanFW,Crevits2019RobustEO}, they present neither algorithmic solutions nor specific designs for our concerned self-driving context. Moreover, we also notice there are some other MOT designs which adopt other state estimation algorithms, such as particle filter \cite{He2016PreciseAE} or probabilistic graphical models \cite{Pang2021MultiObjectTU}. To the best of our knowledge, these alternative designs do not exert as much influence as the KF-based approach on the existing SDC practices, probably due to the additional computing overhead compared with KF. Combined with the above analysis, our current work aims at studying and analyzing the security of the mainstream KF-based MOT and enhances its robustness by a simple yet effective security patch on the original KF mechanism, which, according to the timing experiments in Appendix \ref{sec:app:more_evaluation} on our prototypical implementation, remains over the $24$ fps real-time bar mostly and has almost no negative effects on the normal MOT performance.

 \noindent$\bullet$\textbf{ Limitations and Future Works.} Although the vulnerability of the end-to-end self-driving system against MOT hijacking attacks is revealed in the LGSVL simulator, our proposed security patch is not integrated in such a system for more systematic evaluation, mainly due to the lack of essential data supports for conducting large-scale simulation-based tests \cite{simulation}. As an alternative, we have carefully reimplemented the 2D and 3D MOT module in Baidu's Apollo as an independent runnable codebase, and evaluated the impact of our proposed security patch on the robustness and the performance of the MOT module with real-world benchmark datasets. We think the adopted evaluation setting would reflect the performance of the proposed solution in practical usages to some extension. Nevertheless, considering the generality and the simplicity of our methodology, future works may consider cooperation with the industry to install and test the security patch in both simulation-based and closed-road test cases. 
 
 Moreover, following \cite{Jia2020FoolingDA}, our threat model mainly considers an attacker who has the ability to manipulate the bounding box of the target object by shifting or hiding. Our generalized attack model relaxes the assumption on the arrangement of hiding and shifting frames in the full driving process, under which the robustness of our security patch is validated. Nevertheless, a future attacker who adopts more stronger threat model (e.g., controlling the bounding boxes of multiple objects and combo with bounding box appearing) may find a solution to bypass the proposed security patch and cause a substantial trajectory deviation. However, the realization of stronger threat model incurs more attack requirements and costs. In fact, even in the current threat model, to shift or hide one observed bounding box arbitrarily in consecutive frames requires at least two different adversarial patches generated during the attack phase. Considering the real-time requirement on the attack, the current MOT hijacking model is sufficiently strong for the known attack techniques.

\section{Related Work}
\noindent$\bullet$\textbf{ Attacks and Defenses on the Perception Module.} Self-driving systems adopt multiple sensors and the corresponding deep learning models to perceive the surrounding environments. Consequently, the security of the perception module arouses increasing research interests in the past few years. On the camera-based detection, a number of adversarial patch generation algorithms are proposed to realize object appearing and disappearing attacks \cite{Eykholt2018PhysicalAE,Chen2018RobustPA,Zhao2019SeeingIB}. Recently, Ji et al. also exploit the acoustic signal to perturb the camera-based object detector \cite{Ji2021PoltergeistAA}, while, at the defender's side, Xiang et al. present a certifiably robust object detection model \cite{Xiang2021DetectorGuardPS}. On LiDAR-based detection, Cao et al. propose the first appearing attack by adversarially manipulating the LiDAR point cloud \cite{Cao2019AdversarialOA}, which is later extended to a black-box and more transferable attack in \cite{Sun2020TowardsRL}. Combining with 3D printing techniques, recent research also reveal the vulnerability of existing 3D detection models against object hiding \cite{Tu2020PhysicallyRA,Zhu2021CanWU,Cao2021InvisibleFB} and bounding box enlarging \cite{kaichen2021robust}. At the defender's side, existing approaches are specially designed for LiDAR spoofing based adversarial attacks by exploiting the limitation of physical attack apparatus \cite{Sun2020TowardsRL,hang2019dupnet,radu2008towards,Hau2021ShadowCatcherLI}. 

Different from the attack and defenses on the first half of the perception pipeline, our current work focuses on studying the security of the object tracking module. Adopting the sensor hijacking attack as a prerequisite exploitation techniques, the vulnerability of KF-base MOT is recently demonstrated by Jia et al. \cite{Jia2020FoolingDA} in the context of SDC and by few other works in the context of surveillance cameras \cite{Ding2021TowardsUP,Chen2021AUM}. Our work presents the first certifiably robust security patch which mitigates the effect of sensor hijacking attacks on existing KF-based MOT implementation in self-driving systems.

\noindent$\bullet$\textbf{ Other Security Aspects on SDC.} Besides the progress in understanding the security of the perception pipeline, previous works also extend the security study to the lane detection module \cite{Sato2021DirtyRC,Jing2021TooGT}, the localization module \cite{Wang2021ICS,Shen2020DriftWD,Luo2020StealthyTO}, the planning module \cite{Wan2022TooAT} and the software/hardware levels \cite{Nassi2020PhantomOT,Shin2017IllusionAD,Garcia2020ACS}. Recently, Shen et al. present a comprehensive survey and a unified system-driven evaluation platform on the semantic AI security of self-driving systems \cite{Shen2022SoKOT}.  
\section{Conclusion}
In this paper, we systematize the attack taxonomy of previous MOT hijacking attacks and confirm the known vulnerability of camera-based MOT modules is also present in more complicated MOT implementation of the Apollo system. By carefully inspecting the root cause of this severe vulnerability, we finally discover it is the role inconsistency of the classical KF algorithm in handling the unexpected prediction errors (including adversarial manipulation) from the detection modules that leads to the vulnerability. To address the over-confidence of the existing KF-based MOT implementation in observational data with large deviation, we present a simple yet effective security patch which leverages the modeling of historical observation-prediction deviation to constrain the influence of the abnormal deviation to a safe level. We establish analytical upper bounds on the trajectory deviation with the defense and conduct extensive evaluation to validate the mitigation effectiveness and the trivial performance overhead of our approach. We hope this work would shed light on future development and security enhancing of MOT algorithms in self-driving systems.      

\clearpage
\bibliographystyle{plain}
\bibliography{ref}
\appendix 
\section{More Backgrounds on Kalman Filter}
\label{sec:app:kalman_filter}
Kalman filter (KF) is mainly used in Apollo-MOT to estimating the location-related states. According to the open-sourced implementation \cite{Apollo}, the 2D state $s_{2D}$ contains $(x, v_x, y, v_y)$ and the 3D state $s_{3D}$ contains $(x, v_x, y, v_y, z, v_z)$, corresponding to the observational data $z_{2D} = (x,y)$ and $z_{3D} = (x,y,z)$ (i.e., the center position of the 2D/3D bounding box). For the notation simplicity, we present the backgrounds on KF in the context of the 2D case. In Apollo-MOT, the following constant velocity dynamic model is employed:
\begin{align}
    s_{2D}(t) = As_{2D}(t-1) + v_t \\
    z_{2D}(t) = Hs_{2D}(t-1) + w_t 
\end{align}
where $v_t \sim \mathcal{N}(0, Q)$, $w_t \sim \mathcal{N}(0, R)$ are the systematic and the observational Gaussian noises (the covariance matrices $Q, R$ are hyperparameters), and the state transition operator $A$ and the projection operator $H$ are defined as
\begin{align}
    A = \begin{pmatrix}
    1 & 1 & 0 & 0 \\
    0 & 1 & 0 & 0 \\
    0 & 0 & 1 & 1 \\ 
    0 & 0 & 0 & 1 
    \end{pmatrix}, \quad H = \begin{pmatrix}
    1 & 0 & 0 & 0 \\ 
    0 & 0 & 1 & 0 
    \end{pmatrix}.
\end{align}

Under the above dynamic model, the implemented KF algorithm recursively applies the following prediction (Step 1-2) and updating rules (Step 3-5) to construct the linear minimum variance estimator for the future states. 
\begin{itemize}
    \item \textbf{Step 1.} Predict the next state: $s_{2D}^{-}(t) = As_{2D}(t-1)$.
    \item \textbf{Step 2.} Project the error variance ahead: $P^{-}(t) = AP(t-1)A^{T} + Q$. (Note: $P(0)$ is usually set as the same scale as the covariance matrix of $w_t$.)
    \item \textbf{Step 3.} Compute the Kalman gain: $K(t) = P^{-}(t)H^T(HP^{-}(t)H^T + R)^{-1}$.
    \item \textbf{Step 4.} Merge the observation and the prediction: $ s_\text{2D}(t) = s^{-}_{2D}(t) + K(t)(z_{2D}(t) - Hs_{2D}^{-}(t))$.
    \item \textbf{Step 5.} Update the error covariance: $P(t) = (I - K(t)H)P^{-}(t)$.
\end{itemize}
Specifically, when the observation $z_{2D}(t)$ is missing because the track has no best-matching bounding box, only the first two steps in the above procedure are invoked and the KF algorithm would adopt the predicted state as the final estimated state, i.e., $s_\text{2D}(t) = s^{-}_{2D}(t)$.

\section{More Technical Details in Apollo-MOT}
\label{sec:app:apollo_mot}
We present more technical details on how the MOT module in Apollo conducts the data association process.

\noindent$\bullet$\textbf{ The 2D Data Association Process.} First, Apollo-MOT calculates the similarity score between each pair of trajectories and the observational bounding box, which consists of $4$ similarity dimensions:
\begin{itemize}[topsep=4pt,itemsep=4pt,partopsep=4pt, parsep=2pt]
    \item \textit{Feature Similarity (FS):} FS calculates the dot product between the DNN features of the observed bounding box and the latest bounding box associated with the trajectory. 
    \item \textit{Momentum Similarity (MS):} MS calculates the probability of the predicted state's center $(x', y')$ in the Gaussian distribution derived from the observed bounding box $(x, y, l, w)$, i.e., $MS=\mathcal{N}(x'|x,l)*\mathcal{N}(y'|y,w)$.
    \item \textit{Shape Similarity (SS):} SS measures the relative changes in the bounding box shape from the predicted bounding box of size $L\times{W}$ and the observed bounding box of size $l\times{w}$, i.e., $SS=-|(L-l)*(W-w)/(L*W)|$.
    \item \textit{Intersection Similarity (IS):} IS measures the IoU between the predicted and observed bounding boxes. 
\end{itemize}
The unified similarity score is calculated 
by $0.45*FS+0.4*MS+0.15*SS+0.05*IS$ when the feature similarity is available and by $0.5*MS+0.15*SS+0.35*IS$ otherwise. Next, the MOT algorithm eliminates the candidate pairs which have $MS < 0.045$ or the unified similarity score smaller than $0.6$. Finally, for each trajectory, the MOT algorithm greedily matches the remaining tracks and observed bounding boxes with the negative similarity scores as the pairwise costs.

\noindent$\bullet$\textbf{ The 3D Data Association Process} More comprehensive than the 2D data association, the 3D counterpart calculates the dissimilarity score between each pair of trajectories and the 3D observational bounding box in $7$ dimensions:
\begin{itemize}[topsep=4pt,itemsep=4pt,partopsep=4pt, parsep=2pt]
    \item \textit{Location Dissimilarity (LD)}: LD measures the L2 distance between the centers of the predicted and the observed bounding boxes in the x-y plane. Specifically, if the predicted velocity is larger than $2m/\text{frame}$, the L2 distance is weighted as $(0.5*dx^2+2*dy^2)^{1/2}$.
    \item \textit{Directional Dissimilarity (DD)}: DD measures the cosine value of the angle between the observed bounding box and the latest bounding box associated with the trajectory in the x-y plane.
    \item \textit{Shape Dissimilarity (SD)}: SD measures the relative changes in the bounding box shape from the predicted bounding box of size $L\times{W}\times{H}$ and the observed bounding box of size $l\times{w}\times{h}$, i.e., $SD = |(L-l)*(W-w)*(H-h)/(L*W*H)|$.
    \item \textit{Point Density Dissimilarity (PDD):} PDD measures the relative changes in the density of the point cloud covered in the observed and the latest associated bounding boxes, i.e., $PDD=|n-m|/max(n, m)$ ($n$, $m$ is the number of points in the point clouds).
    \item \textit{Feature Dissimilarity (FD)}: FD measures the L1 distance between the DNN features of the observed and the latest associated bounding box. For the cases when the feature dimensions are inconsistent with one another, FD is set as $100$. 
    \item \textit{Mass Center Dissimilarity (MCD)}: MCD calculates the L2 distance between the mass centers (in terms of the covered point cloud) of the observed and the latest associated bounding box. 
    \item \textit{Intersection Dissimilarity (ID)}: ID measures the 1-IoU of the predicted and the observed bounding boxes in the $x-y$ plane, which is hard-thresholded by a matching constant.   
\end{itemize}
When the observed object is predicted to be in the foreground, the unified dissimilarity score is calculated as $0.6*LD+0.2*DD+0.1*SD+0.1*PDD+0.5*FD$. Otherwise, the dissimilarity score is calculated as $0.2*MCD+0.8*ID$. Next, the MOT algorithm eliminates the candidate matches which has a dissimilarity score larger than $4.0$. Finally, the trajectory and the observed bounding boxes are matched with the Hungarian algorithm using the dissimilarity scores as the pairwise costs.

\section{Omitted Technical Proofs}
\label{sec:app:proofs}
\noindent$\bullet$\textit{ Setups.} In this part, we provide the omitted proofs for our analytical results in Section \ref{sec:security_analysis}. Specifically, our results establish the upper bound on the trajectory deviation under the generalized MOT hijacking attack when our security patch is deployed or not. Without loss of generality, we mainly study the deviation at the horizontal axis where the attacker targets, while similar deviation bound can be obtained for other axis. First, we rewrite different types of MOT updating rules in the continuous-time form. We use $s_{ij}$ to denote the state variable in different scenarios, where $i, j\in\{0,1\}$ denote whether the observation is hijacked and whether the defense is deployed respectively. 

 The original KF is rewritten as
\begin{equation}
    d{s}_{00} = As_{00}dt + K(dz - Cs_{00}dt),
\label{eq:original}
\end{equation}
where $C := HA$ and $dz (:= Hds + dw_t)$ denotes the raw observational data without malicious manipulation which satisfies the light-tailed assumption that $P\{|dw_t| > \delta \} < e^{-\delta}$. The differential equation above can be compared with the discrete-form updating rule in Eq.(\ref{eq:kf_merging}). According to \cite{}, the continuous-time transform would have trivial errors on deriving results on the original form. 

When the observational data is hijacked to be $dz' (:= Hds + dw_t')$ such that the deviation $dw_t'$ satisfies $dw_t' = dw_t + \lambda s_t dt$ ($s_t$ is a Bernoulli random variable of probability $s\%$ to be $1$, indicating a shifting attack frame), the original KF updating equation is perturbed to be
\begin{equation}
        ds_{10} = As_{10}dt + h_tK(dz' -Cs_{10}dt), 
\label{eq:ori_kf_with_atk}
\end{equation}
where $h_t$ is a Bernoulli random variable of probability $r\%$ to be $1$, indicating a hiding attack frame. To model the exclusivity of the two types of attack frames, we have $P(s_t=0|h_t=1)=1$ and $P(h_t=0|s_t=1)=1$.

Similarly, when our defense is equipped, the KF updating equation with or without hijacking is written as $d{s}_{01} = As_{01}dt + K\phi(dz - Cs_{01}dt)$ and $d{s}_{11} = As_{11}dt + Kh_t\phi(dz' - Cs_{11}dt)$. To understand how the deviation in different attack/defense configurations accumulates along the driving, we inspect the evolutionary equation for $\Delta_{ij} := |s_{ij} - s_{00}|$. 

\begin{proof}[Proof for Proposition \ref{prop:attack_effectiveness}]
First, for $i = 1$ and $j = 0$, we derive the following equation by subtracting Eq.(\ref{eq:ori_kf_with_atk}) from both sides of Eq.(\ref{eq:original}):
\begin{equation}
\begin{cases}
    d\Delta_{10} = (A-KC)\Delta_{10}{dt} + K(dw_t - h_tdw_t') \\ \Delta_{10}(0) = 0
\end{cases}
\label{eq:delta_10_evolution}
\end{equation}
where $\Delta_{10}(t) = s_{10}(t) - s_{00}(t)$ and the initial condition $\Delta_{10}(0) = 0$ is because the initial state value for $s_{10}$ and $s_{00}$ is identical. Constructing the state operation $\Psi(t)$ which satisfies $\dot{\Psi}(t) = (A-KC)\Psi(t)$, we have for every $t > s > 0$, there exists a positive constants $\beta$ such that 
\begin{equation}
    \|\Psi(t)\Psi(s)^{-1}\| \sim \Theta(e^{-\beta(t-s)}),
\end{equation}
where $f(t) \sim g(t) $ denotes $f(t)$ grows in the same scale as $g(t)$ and the constant $\beta$ is related with the eigenvalues of the operator $A - KC$. 


Therefore, by integrating Eq.(\ref{eq:delta_10_evolution}) from $t=0$ to $t=T$ at both sides, we have 
\begin{align}
    & \Delta_{10}(T) = \int_{0}^{T}\Psi(T)\Psi(s)^{-1}K(dw_s - h_sdw_s^{'})ds \nonumber \\
    &  = \int_{0}^{T}\Theta(e^{-\beta(T-s)})K(dw_s - h_sdw_s^{'}) \\
    & \overset{\|K\|_\infty < \infty}{\sim} \int_{\mathcal{T}_s}\lambda\Theta(e^{-\beta(T-s)})ds + \int_{\mathcal{T}_h}\Theta(e^{-\beta(T-s)})dw_s
\end{align}
where the first term measures the trajectory deviation during the shifting frames (i.e., the set of timestamps $\mathcal{T}_s$ s.t. $P(\mathcal{T}_s) = s\%$), and the second term measures the deviation during the hiding frames (with $P(\mathcal{T}_h) = h\%$). We notice that the first term has no dependence on where the shifting frames $\mathcal{T}_s$ are distributed in the total driving frames, while the second term, as an Ito integral over the Brownian motion defined by $dw_s$ \cite{oksendal2013stochastic}, is sensitive to the initial value at the first hiding frame. Therefore, arranging the shifting frames $\mathcal{T}_s$ immediately before the hiding frames $\mathcal{T}_h$, i.e., $\mathcal{T}_s = [(1-r\%)T, (1-h\%)T]$ and $\mathcal{T}_h = ((1-h\%)T, T]$, would result in the maximal deviation. By directly calculating the first integral, we have $\int_{\mathcal{T}_s}\lambda\Theta(e^{-\beta(T-s)})ds = \frac{\lambda}{\beta}\Theta(e^{-\beta s\% T})$. Involving the derived deviation as the initial value of the Ito process in the second term, we derive the upper bound in Proposition \ref{prop:attack_effectiveness}. 
\end{proof}

\begin{proof}[Proof for Proposition \ref{prop:defense_overhead}]
Following the similar procedures as above, we derive the explicit formula for the deviation growth for $s_{01}$ and $s_{11}$. For $s_{01}$, we have
\begin{equation}
\Delta_{01}(T) \sim \int_{\{|dw_t| > \delta_\text{max}\}}\Theta(e^{-\beta(T-s)})(dw_t - \delta_\text{max}) {dt}. 
\end{equation}
As the benign observation $dw_t$ follows a light-tailed distribution and $\delta_\text{max}$ is chosen as the $95\%$ quantile of the normal observation distribution, the above integral would mainly happens for a discrete set of frames when the dynamic model underlying MOT characterizes less accurately about the target object (i.e., $\{|dw_t| > \delta_\text{max}\}$), which vanishes because the set of $\{|dw_t| > \delta_\text{max}\}$ has a zero measure \cite{walter1974real}. 
 
\end{proof}

\begin{proof}[Proof for Proposition \ref{prop:defense_effectiveness}]
For $s_{11}$, we have 
\begin{align}
    &\Delta_{11}(T) 
    \sim \int_{\{|dw_s'| < \delta_\text{max}\}\cap{\mathcal{T}_s}}\lambda\Theta(e^{-\beta(T-s)})ds \nonumber \\ 
    &+ \int_{\{|dw_s'| \ge \delta_\text{max}\} \cap {{\mathcal{T}_s}}}\Theta(e^{-\beta(T-s)})(\delta_\text{max}ds - dw_s)  \nonumber \\ 
    & + \int_{\mathcal{T}_h}\Theta(e^{-\beta(T-s)})dw_s.
\end{align}
The second term vanishes due to the same arguments in our proof for Proposition \ref{prop:defense_overhead}. With the similar analysis in our proofs for Proposition \ref{prop:attack_effectiveness}, we note that the deviation is maximized when $\mathcal{T}_s = [(1-r\%)T, (1-h\%)T]$ and $\mathcal{T}_h = ((1-h\%)T, T]$. Specifically, we calculate the integral in the first term, we obtain that the maximal deviation during the shifting phase is $\frac{\delta_\text{max}}{\beta}\Theta(e^{-\beta s\% T})$, and use the deviation as the start point in the stochastic integral in the third term, which gives the final upper bound in Proposition \ref{prop:defense_effectiveness}.  
\end{proof}

\section{More Evaluation Results}
\label{sec:app:more_evaluation}
We present a tentative timing experiment on the Apollo 3DMOT with or without our proposed security patch (the implementation of which is prototypical without additional efficiency optimization). Specifically, we run the MOT algorithm (without the object detection module) in the same experimental environment on  $5$ randomly sampled traces from KITTI. On each trace, we run the timing for $5$ repetitive times and reports the average running time with the corresponding FPS in Table \ref{tab:timing}.

\begin{table}[t]
  \centering
  \caption{A tentative timing comparison on Apollo 3DMOT with or without our proposed security patch.}
  \scalebox{0.9}{
    \begin{tabular}{lrrrrr}
    \toprule
          &       & \multicolumn{2}{c}{w/o. Defense} & \multicolumn{2}{c}{w/. Defense} \\
    \cmidrule(lr){3-4}  \cmidrule(lr){5-6} 
          & \multicolumn{1}{c}{\# Frames} & \multicolumn{1}{c}{Time (\textit{sec.})} & \multicolumn{1}{c}{FPS} & \multicolumn{1}{c}{Time (\textit{sec.})} & \multicolumn{1}{c}{FPS} \\
    \midrule
    Trace \#1 & 145   & 0.17  & 842   & 2.24  & 64.8 \\
    Trace \#2 & 376   & 1.34  & 280   & 6.68  & 22.7 \\
    Trace \#3 & 340   & 0.84  & 406   & 8.1   & 42 \\
    Trace \#4 & 78    & 0.14  & 561.1 & 2.38  & 32.7 \\
    Trace \#5 & 270   & 0.6   & 453.2 & 9.28  & 29.1 \\
    \bottomrule
    \end{tabular}}%
  \label{tab:timing}%
\end{table}%

\noindent\textbf{Experimental Environments.} The Apollo 3DMOT in the timing experiments is carefully reimplemented with Python according to the official implementation in \cite{Apollo} and the timing results are collected on a Linux server running Ubuntu 16.04, 64GB memor, one AMD Ryzen Threadripper 2990WX 32-core processor and 2 NVIDIA GTX RTX2080 GPUs.

\end{document}